\documentclass[colorlinks,citecolor=blue,urlcolor=magenta,linkcolor=blue]{article}
\usepackage{graphicx}  
\usepackage{amsmath}   
\usepackage[compress,numbers,sort]{natbib}
\usepackage{amssymb}   
\usepackage{bm} 
\usepackage{dcolumn}
\usepackage{color}
\usepackage{mathrsfs}
\usepackage{float}
\usepackage{amsfonts}
\usepackage{varioref}
\usepackage{textcomp}
\usepackage[normalem]{ulem}

\usepackage{multirow}
\usepackage{caption}
\usepackage{orcidlink}
\usepackage{subcaption}
\allowdisplaybreaks
\labelformat{section}{Section #1} 
\labelformat{subsection}{Section #1} 
\labelformat{subsubsection}{Section #1}
\labelformat{subsubsubsection}{Section #1}
\labelformat{equation}{Eq.~(#1)} 
\labelformat{figure}{Fig.~#1} 
\labelformat{subfigure}{Fig.~\thefigure#1} 
\labelformat{table}{Table~#1} 
\labelformat{appendix}{Appendix #1}

\begin{document}

\author{Surojit Dalui \orcidlink{0000-0003-1003-8451} \footnote{surojitdalui@shu.edu.cn, surojitdalui003@gmail.com}$~^{1}$, Chiranjeeb Singha \orcidlink{0000-0003-0441-318X}\footnote{chiranjeeb.singha@iucaa.in}$~^{2}$, and Krishnakanta Bhattacharya\orcidlink{0000-0003-3309-2610}
\footnote{krishnakanta@dubai.bits-pilani.ac.in}$~^{3}$
\\$^{1}${\small{Department of Physics, Shanghai University, 99 Shangda Road, Baoshan District,}}\\
{\small{  Shanghai 200444, People's Republic of China}}
\\$^{2}${\small{Inter-University Centre for Astronomy and Astrophysics (IUCAA), Post Bag 4, }}\\
{\small{ Ganeshkhind, Pune 411007, India}}
\\$^{3}${\small{Department of General Science, Birla Institute of Technology and Science, Pilani,}}\\
{\small{Dubai Campus, International Academic City, Dubai, United Arab Emirates}}\\}
 
\title{\bf Chaotic Dynamics in Extremal Black Holes: A Challenge to the Chaos Bound}
\date{\today}

\maketitle

\begin{abstract} 
\noindent
We investigate chaotic dynamics in extremal black holes by analyzing the motion of massless particles in both Reissner-Nordstr\"{o}m and Kerr geometries. Two complementary approaches (i) taking the extremal limit of non-extremal solutions and (ii) working directly in the extremal background, yield consistent results. We find that, contrary to naive extrapolation of the Maldacena-Shenker-Stanford (MSS) chaos bound, the Lyapunov exponent remains positive even at zero temperature. For Reissner-Nordstr\"{o}m black holes, chaos diminishes but persists at extremality, while for Kerr black holes it strengthens with increasing spin. These results demonstrate that extremal black holes exhibit residual chaotic dynamics that violate the MSS bound, establishing them as qualitatively distinct dynamical phases of gravity.

\end{abstract}
\section{Introduction}
Black holes are considered remarkable objects in theoretical physics for their extraordinary classical properties and for providing deep insights into quantum gravity. Out of the different black holes that exist in literature, the extremal black holes, which are characterized by maximal charge or spin, are particularly interesting as compared to their nonextremal counterparts for several reasons. For example, from the viewpoint of thermodynamics, extremal black holes show a fascinating and unusual trait: even though their Hawking temperature drops to zero, they still have a finite amount of entropy. This is in agreement with the statistical mechanics interpretation. Moreover, recent theoretical investigations have intensified the debate surrounding extremal black holes, questioning their fundamental nature and relation to nonextremal black holes. A critical point of this ongoing discussion is whether extremal black holes should be viewed as separate entities with unique dynamical and thermodynamic identities or as merely smooth limiting cases of nonextremal black holes. Recent studies suggest \cite{Nanda:2022szu, Ghosh:1995rv} that extremal black holes could be a unique phase of matter, standing out because of their strange thermodynamic behavior. These ongoing discussions emphasize how important it is to conduct deeper investigations into the behavior of extremal black holes and what they can teach us about gravity.

While much of the earlier work has addressed these issues from the perspective of geometry \cite{Carroll:2009maa, Pradhan:2012yx, Howard:2013yqq,Gibbons:1994ff}, symmetry \cite{Guica:2008mu,Castro:2010fd}, and quantum microstates \cite{Strominger:1996sh,Lunin:2001jy,Sen:2008vm}, fewer studies have explored the dynamical implications of extremality. In this regard, the recent developments \cite{Bombelli:1991eg, Sota:1995ms,Vieira:1996zf,Suzuki:1996gm,Cornish:1996ri,deMoura:1999wf,Hartl:2002ig,Han:2008zzf,Takahashi:2008zh,Li:2018wtz,Hashimoto:2016dfz,Dalui:2018,Dalui:2019,Bera_2022,Das_2024,Giataganas:2021ghs} in the study of chaotic behavior near black hole horizons may provide a novel window into this topic. In recent years, chaos has become an important and revealing tool in gravitational physics, especially when it comes to exploring what happens near the horizons of black holes. Researchers have been studying chaotic behavior in these regions to understand the fundamental nature of gravity, the fabric of spacetime, and how these ideas connect to quantum theories of gravity. The role of chaos, quantified through Lyapunov exponents, is particularly significant in probing the properties and characteristics of extremal black holes. Several studies have established the connection between classical chaos indicators and quantum field theoretic outcomes, further underscoring chaos as a bridge between classical and quantum regimes in black hole physics \cite{Dalui_2020, Dalui_2020_1,Dalui_2022}. A particularly important aspect of this connection is embodied in the Maldacena–Shenker–Stanford (MSS) chaos bound, which relates the maximal Lyapunov exponent to the temperature of quantum chaotic systems as $\lambda_L\leq 2\pi T/\hbar$ \cite{Maldacena_2016}. This bound has been shown to hold for a broad class of nonextremal black holes \cite{Hashimoto:2016dfz,Dalui:2018,Das_2024}, thereby providing a stringent test for their near-horizon dynamics and quantum stability. Extremal black holes, however, remain a special case. Despite having finite entropy, their Hawking temperature vanishes, and naive extrapolation of the MSS bound would imply the absence of chaos. At the same time, extremal geometries can be regarded either as smooth limits of non-extremal solutions or as distinct configurations with unique near-horizon structure. Whether chaotic dynamics persist at extremality, and whether they can be captured as a limiting case of near-extremal black holes, remains an open question of both conceptual and practical importance.


In this study, we aim to address these fundamental questions by studying the chaotic behavior of test particles in strong gravitational fields by analyzing the motion of a massless, chargeless particle in the near-horizon region of black holes. Our analysis encompasses both rotating (Kerr) and non-rotating but electrically charged (Reissner–Nordström, RN) black hole geometries, thereby allowing us to disentangle the roles of angular momentum and charge in shaping the dynamics. We adopt two complementary methodological approaches to probe the interplay between extremal and nonextremal configurations.
In the first approach, we consider nonextremal black holes and gradually tune their parameters, specifically, the spin in the Kerr case and the electric charge in the RN case, toward the extremal limit. This procedure enables us to examine how the dynamical features evolve as the system approaches extremality, thereby treating the extremal state as a limiting case of continuous deformation. In contrast, the second approach treats the extremal black hole background as a distinct configuration with unique near-horizon geometry, rather than as a mere endpoint of a parametric sequence. Within this framework, we directly study the motion of particles in the extremal spacetime and contrast it with the nonextremal case.
We perform detailed numerical simulations for both approaches to compute particle trajectories and extract quantitative measures of dynamical instability. In particular, we evaluate the Lyapunov exponents associated with radial perturbations of geodesic motion. These exponents serve as precise indicators of sensitivity to initial conditions and are widely regarded as diagnostic tools for identifying chaotic behavior in dynamical systems. Our analysis shows that, while the two approaches differ in perspective, one emphasizing the continuity of the extremal limit and the other highlighting the distinctiveness of extremality, the resulting dynamical features exhibit strong consistency. In both cases, the computed Lyapunov exponents demonstrate qualitatively similar behavior, with the growth rates saturating at finite values that remain well below the universal chaos bound proposed by Maldacena, Shenker, and Stanford (MSS). This outcome suggests that, although extremal black holes exhibit intriguing dynamical features, their chaotic behavior remains constrained in a fundamentally distinct way from the maximal chaoticity allowed by the MSS bound. These results establish that extremal black holes represent a distinct dynamical phase, characterized by residual chaos beyond the MSS bound, and highlight their special role in the broader landscape of gravitational and quantum chaotic systems.

The paper is organized as follows. In the next section, we briefly overview the spherically symmetric Reissner–Nordstr\"{o}m black hole and the rotating Kerr black hole, emphasizing the characteristics of their respective extremal limits. We highlight the distinctive thermodynamic properties of extremal black holes compared to their nonextremal counterparts, and review the ongoing debate in the literature regarding whether extremal black holes should be considered as limiting cases or as fundamentally different objects. In \ref{sec3}, we introduce the general framework for analyzing particle dynamics by considering a massless particle propagating in the Reissner–Nordstr\"{o}m and Kerr black hole spacetimes. \ref{sec4} is dedicated to the computation of the Lyapunov exponent for these backgrounds. As outlined earlier, we pursue this analysis via two complementary approaches: (i) by first computing the result for nonextremal black holes and subsequently taking the extremal limit, and (ii) by directly starting from the extremal black hole geometry. Finally, we summarize our findings and discuss future directions in \ref{conclusion}.

\textbf{\textit{Notations and Conventions}}:  
Throughout this paper, we adopt the mostly positive signature convention, where the Minkowski metric in $1+3$ dimensions is given in Cartesian coordinates as $\text{diag}(-1, +1,\\ +1, +1)$. Furthermore, we utilized the geometrized units throughout the paper where $G = c = 1$.

\section{Representative Black Holes and the Question of Continuity in Extremal Limits} \label{sec2}

In this section, we examine two notable black hole solutions that become extremal in a specific limit. One is spherically symmetric, and the other is axisymmetric and rotating. Both are among the most well-known black holes in general relativity.

\subsection{Reissner-Nordstr\"{o}m black hole}

The spherically symmetric black hole is considered here as a Reissner–Nordstr\"{o}m black hole, described by the metric \cite{reissner_h_1916_1447315,book:15209}, 
\begin{equation}
ds^2=-f(r)dt^2+\frac{1}{f(r)}dr^2+r^2d\Omega^2~,\label{RN metric} ~~~~~~~~~~~~~~\textrm{with}~~~~~~~~~~~~~~~~~~
f(r) = \left(1 - \frac{2M}{r} + \frac{Q^2}{r^2}\right)
\end{equation}
and $M$ and $Q$ are the mass and electric charge of the black hole. The horizon radius is determined by the conditions $f(r) = 0$, which yields two solutions:  the outer ($r_+$) and the inner horizon ($r_-$) that are given as
\begin{equation}
r_{\pm} = M \pm \sqrt{M^2 - Q^2}~.
\end{equation}
The outer horizon is the event horizon, and the inner horizon is the Cauchy horizon. Since the Reissner-Nordstr\"{o}m metric is spherically symmetric, the surface gravity for the event horizon can be obtained conventionally as, 
\begin{equation}
\kappa = \frac{1}{2} f^{\prime} |_{r=r_+} = \frac{\sqrt{M^2 - Q^2}}{r_+^2}=\frac{r_+-r_-}{2r_+^2}~.
\end{equation}
The entropy of the black hole, following the Bekenstein–Hawking area law, is:
\begin{align}
    S=\pi r_+^2=\pi\bigg(M+\sqrt{M^2-Q^2}\bigg)^{2}~.
\end{align}
The extremal limit of the Reissner–Nordström black hole is reached when $Q=M$. The metric for an extremal black hole is given by,
\begin{equation}
ds^2=-f(r)dt^2+\frac{1}{f(r)}dr^2+r^2d\Omega^2~,\label{RN etremal metric} ~~~~~~~~~~~~~~\textrm{with}~~~~~~~~~~~~~~~~~~
f(r) = \left(1 - \frac{M}{r}\right)^{2}~.
\end{equation}
In this case:
\begin{itemize}
    \item The two horizons for a Reissner–Nordström black hole coincide, that is, $r_+=r_-=M$.
    \item The surface gravity, or the Hawking temperature, vanishes.
    \item Yet, the entropy remains finite and non-zero: $S=\pi M^2=\pi r_+^2$.
\end{itemize}
The extremal case lies on the boundary between physical black holes and naked singularities, and represents a highly symmetric and stable configuration in both classical and quantum contexts.
\subsection{Kerr black hole}
The axisymmetric and rotating black hole is considered here as a Kerr black hole, described by the metric in Boyer-Lindquist coordinates \cite{Kerr:2007dk, Teukolsky:2014vca, Visser:2007fj},
\begin{align}
ds^2 = -\left(1 - \frac{2Mr}{\Sigma}\right) dt^2 
- \frac{4Mar\sin^2\theta}{\Sigma} dt\, d\phi 
+ \frac{\Sigma}{\Delta} dr^2 
+ \Sigma\, d\theta^2 
+ \left(r^2 + a^2 + \frac{2Ma^2r\sin^2\theta}{\Sigma}\right)\sin^2\theta\, d\phi^2\label{kerr_metric}
\end{align}
where 
\begin{align}
    \Sigma = r^2 + a^2\cos^2\theta, \qquad\qquad\ \Delta = r^2 - 2Mr + a^2~.
\end{align}
Here $M$ is the mass of the black hole, $a=J/M$ is the spin parameter (angular momentum per unit mass). The horizons  (outer, inner) are determined by $\Delta = 0$ and the horizons (event horizon $r_+$ and the inner Cauchy horizon $r_-$) are given as 
\begin{align}
    r_{\pm}=M \pm \sqrt{M^2 - a^2}\label{RN_horizon}~.
\end{align}
The Hawking temperature of the Kerr black hole is given as
\begin{align}
    T=\frac{\kappa}{2\pi}~~~~~~~~~~~~~~~~\textrm{where}~~~~~~~~~~~~~~~~\kappa=\frac{r_+-r_-}{2(r_+^2+a^2)}~.
\end{align}
The entropy of the black hole is given as
\begin{align}
    S=\pi (r_+^2+a^2)~.
\end{align}
In the extremal limit, where $a\rightarrow M$, one finds that the two horizons coincide: $r_+=r_-=M$. The spacetime metric of the Kerr black hole in the extremal limit is given as
\begin{align} \label{kerr_extremal}
    ds^2 &= -\left(1 - \frac{2Mr}{\Sigma} \right) dt^2 - \frac{4M^2 r \sin^2\theta}{\Sigma} \, dt \, d\phi + \frac{\Sigma}{(r - M)^2} \, dr^2 \nonumber\\
    &+ \Sigma \, d\theta^2 + \left( r^2 + M^2 + \frac{2M^3 r \sin^2\theta}{\Sigma} \right) \sin^2\theta \, d\phi^2
\end{align}
with
\begin{align}
    \Sigma=r^2+M^2\cos^2\theta~.
\end{align}
In that case, Hawking temperature again vanishes, yet entropy remains a non-zero vanishing quantity $S=2\pi M^2$. Thus, as in the Reissner–Nordstr\"{o}m case, the extremal Kerr black hole challenges standard thermodynamic expectations by exhibiting zero temperature with non-zero entropy.
\subsection*{Extremal Black Holes: A testbed for the MSS Bound}
As discussed in the previous subsections, the Hawking temperature vanishes for all extremal black holes, while the entropy remains a finite, non-zero quantity. This apparently contradicts conventional thermodynamic intuition, leading to two contrasting viewpoints in the literature.

\begin{enumerate}
\item The transition from nonextremal to extremal is continuous \cite{Balbinot:2007kr, Galli:2011fq, Andrianopoli:2013kya,Lemos:2010kw,Kiefer:1998rr,Bhattacharya:2019awq}, and extremal black holes can have truly non-zero entropy despite vanishing temperature due to specific underlying microstates. As shown in the seminal work of Strominger and Vafa \cite{Strominger:1996sh}, certain black holes correspond to the bound state of branes that can be attributed to a large degeneracy of BPS microstates, which yields a statistically significant entropy. This provides the explanations why the extremal black holes can have finite entropy, which originates from the underlying microscopic description and is independent of their thermal considerations.

\item The transition from nonextremal to extremal black holes is not continuous. Consequently, extremal black holes should be treated as qualitatively distinct objects, and their thermodynamics should not be obtained merely by taking the extremal limit in the final result, originally obtained for the nonextremal black holes. Very recently, it has been demonstrated that when one carefully analyzes the near-horizon geometry of an extremal black hole, specifically by expanding the metric near the horizon up to second order, it leads to a surprising result: the temperature of the extremal black hole is no longer exactly zero. This finding challenges the conventional wisdom that extremal black holes are strictly zero-temperature objects and suggests a more subtle thermodynamic behavior in the extremal limit \cite{Nanda:2022szu}. This result is also consistent with the previous finding that the Hawking temperature of an extremal black hole is completely arbitrary \cite{Ghosh:1995rv}. In this view, results derived for nonextremal black holes cannot simply be extrapolated to the extremal case by taking the extremal limit. This perspective has substantial support, particularly from topological and thermodynamic analyses \cite{Gibbons:1994ff,Hawking:1994ii,Teitelboim:1994az,Sen:2008vm}. If correct, this implies that the conventional method of deriving temperature and entropy for nonextremal black holes and then taking the extremal limit is invalid.
\end{enumerate}
Given these contrasting viewpoints, we aim to investigate whether the chaotic dynamics of an extremal black hole can be obtained as the limiting case of a non-extremal one. If this proves possible, as we shall demonstrate in the following analysis, then the situation becomes even more intriguing, as it hints at the possibility of obtaining results for extremal BHs via extrapolating the results of a near-extremal BH. As discussed earlier, the Hawking temperature of an extremal black hole vanishes when extrapolated from the non-extremal result. Consequently, if one extrapolates the Maldacena–Shenker–Stanford (MSS) bound to the extremal case, the Lyapunov exponent would be expected to become negative. However, what we find is that, although the chaotic dynamics of the extremal case can indeed be obtained as the limit of near-extremal black holes (with the Lyapunov exponent matching in both cases), the exponent remains positive. Thus, the MSS bound is violated in the extremal limit.



In the following, we present the dynamics of massless particles. Subsequently, we explore their behavior in the backgrounds of Reissner-Nordström and Kerr black hole spacetimes.

\section{Dynamics of a massless particle} \label{sec3}

\subsection{Reissner-Nordstr\"{o}m black holes spacetime} 

In this part, we summarize the dynamic equations of motion of a massless particle in a Reissner-Nordström (RN) black hole spacetime. In order to find the energy of the particle $E=-\zeta^{\alpha}p_{a}$ (where $\zeta^{\alpha}$ is the Killing vector and $p_{\alpha}$) is the particle's four-momentum), we use the covariant form of the dispersion relation given by
\begin{equation}
	g^{\alpha\beta}p_{\alpha}p_{\beta}=-m^2 c^2~,\label{3.1}
\end{equation}
with $m$, being the mass of the particle. Now, using the dispersion relation \ref{3.1} (with $c=1$) for the metric \ref{RN metric}, we can get the energy of a massless particle ($m=0$) as follows (considering the +ve sign solution of $E$): 
\begin{equation}
 E=\sqrt{f(r)\Big[f(r)p_r^2+\dfrac{p_{\theta}^2}{r^2}\Big]}~,\label{3.2}
\end{equation}
where we have assumed the motion of the particle is in the poloidal plane with $p_{\phi}=0$.

Now, let us construct the dynamical equations of motion for a probe massless particle moving in the background geometry of the RN black hole. To prevent particle trajectories from crossing the horizon, we introduce harmonic confinement along the radial ($r$) and angular ($\theta$) directions, characterized by oscillator strengths $K_r$ and $K_{\theta}$, respectively. Adjusting these strengths allows particle trajectories to be confined within any desired finite region. The harmonic oscillators along radial and angular directions represent an external potential chosen explicitly for confinement purposes, and are not derived from any intrinsic physical property of the black hole background. Their role is crucial in this context because they ensure numerical stability, enabling the detailed investigation of particle dynamics near the horizon. Importantly, these harmonic potentials are explicitly static, do not introduce any external energy sources, or impose any time-dependent perturbations. Their sole purpose is to provide a stable equilibrium situation, ensuring the particle remains within a well-defined region for numerical analysis. Thus, any observed chaotic dynamics or instabilities are fundamentally gravitational in origin, driven by the intrinsic properties of the spacetime near the horizon rather than induced artificially.

We thus examine a scenario where a probe massless particle is subjected to two harmonic potentials, given by $\frac{1}{2}K_r(r - r_c)^2$ and $\frac{1}{2}K_\theta(y - y_c)^2$, along the radial and angular directions, respectively. Here, $K_r$ and $K_\theta$ denote the oscillator strengths in the radial and angular directions (while writing the dynamical equations, we introduce a new variable $y = r_{+}\theta$). The equilibrium positions of these harmonic potentials are denoted as $r_c$ and $y_c$. This modeling approach has previously been suggested in the literature (Ref. \cite{Hashimoto:2016dfz} for massive particles, and Refs. \cite{Dalui:2018,Dalui:2019,Bera_2022} for massless particles). It should be emphasized that replacing these harmonic potentials with alternative potentials could modify the particle dynamics. However, as discussed extensively in Ref. \cite{Dalui:2018,Dalui:2019}, for massless particles, the radial motion remains unaffected by changes to the specific form of the potential.

Now the total energy of the probe particle under the influence of harmonic potentials for the metric \eqref{RN metric} is given by
\begin{eqnarray}
E_{\text{RN}}=\sqrt{f(r)\Big[f(r)p_r^2+\dfrac{p_{\theta}^2}{r^2}\Big]}+\frac{1}{2}K_r(r-r_0)^2 +\frac{1}{2}K_\theta~r^2_{+}(\theta-\theta_0)^2~.\label{3.3}
\end{eqnarray}
Let us note that the radial momenta $p_{r}$ and cross radial momenta $p_{\theta}$, which appear in the above energy expression, are the usual canonical momenta that can be derived from the Lagrangian of a massless particle in the Schwarzwald coordinate system.

Correspondingly, the dynamical equations of motion have the following form:
\begin{eqnarray}
	\dot{r}&=&\dfrac{\partial E_{\text{RN}}}{\partial p_r}=\frac{f^2(r)p_r}{\sqrt{f^2(r)p_r^2+\dfrac{f(r)p_{\theta}^2}{r^2}}}~,\label{3.4}
	\\
	\dot{p_r}&=&-\frac{\partial E_{\text{RN}}}{\partial r}=-\dfrac{f(r)f'(r)p_r^2}{\sqrt{f^2(r)p_r^2+\dfrac{f(r)p_{\theta}^2}{r^2}}}-K_r(r-r_0)\nonumber\\
	&&-\dfrac{f'(r)p_{\theta}^2}{2r^2\sqrt{f^2(r)p_r^2+\dfrac{f(r)p_{\theta}^2}{r^2}}}+\dfrac{f(r)p_{\theta}^2}{r^3\sqrt{f^2(r)p_r^2+\dfrac{f(r)p_{\theta}^2}{r^2}}}~,\nonumber\\
	\label{3.5}
	\\
	\dot{\theta}&=&\frac{\partial E_{\text{RN}}}{\partial p_{\theta}}=\dfrac{f(r)p_{\theta}}{r^2\sqrt{f^2(r)p_r^2+\dfrac{f(r)p_{\theta}^2}{r^2}}}~,\label{3.6}
	\\
	\dot{p_{\theta}}&=&-\frac{\partial E_{\text{RN}}}{\partial\theta}=-K_{\theta}~r^2_+(\theta-\theta_0)~.\label{3.7}
\end{eqnarray}
where the derivative is taken with respect to an affine parameter $\tau$. These are the main equations for the studies of Lyapunov exponents in the later sections for the RN black hole.

\subsection{Kerr black hole spacetime } 
In a similar way, we shall summarize the dynamical equations of motion of a massless particle in a Kerr black hole spacetime in this part. Using the dispersion relation, we find the energy of the massless particle in the Kerr background (\ref{kerr_metric}) as (considering the +ve sign solution of $E$):
\begin{equation}
	E = 
	\sqrt{\frac{2\Delta^2 p_r^2 + 2\Delta\Sigma p_{\theta}^2}
		{\Sigma\left[a^4 + 2 r^4 + a^2 r(2 M + 3 r) + a^2 \Delta \cos 2\theta\right]}}~.\label{4.1}
\end{equation}
Here also, we have assumed that the particle is moving only along the radial $r$ and the angular $\theta$ directions, keeping $p_{\phi}=0$. 

Now, introducing the harmonic potentials just like the earlier section, we get the total energy of the system as:
\begin{equation}
	E_{\text{Kerr}}= 
	\sqrt{\frac{2\Delta^2 p_r^2 + 2\Delta\Sigma p_{\theta}^2}
		{\Sigma\left[a^4 + 2 r^4 + a^2 r(2 M + 3 r) + a^2 \Delta \cos 2\theta\right]}}+\frac{1}{2}K_r(r-r_c)^2+\frac{1}{2}K_\theta~r^2_{+}(\theta-\theta_c)^2~.\label{4.2}
\end{equation}
Correspondingly, the dynamical equations of motion are as follows: 
\begin{eqnarray}
	&&\dot{r} = \dfrac{\partial E_{\text{Kerr}}}{\partial p_r}=\frac{\Delta\,p_r \left(a^2 + r^2\right) \left(a^2 + r^2 + a^2\cos2\theta\right)}
	{\Sigma^2 \sqrt{\frac{\Delta\,p_r^2}{\Sigma} + \frac{p_{\theta}^2}{\Sigma^2}
			\frac{\left(a^2 + r^2\right)^2 - a^2\Delta\sin^2\theta}{\sin^2\theta}}}~,\label{4.3}\\
	&&\dot{p_r} =-\dfrac{\partial E_{\text{Kerr}}}{\partial r}= -\frac{1}{2E}\Biggl[
		\frac{\partial}{\partial r}\left(\frac{\Delta}{\Sigma^2}\right)p_r^2 
		+ \frac{\partial}{\partial r}\left(\frac{1}{\Sigma}\right)p_{\theta}^2\nonumber\\
		&&~~~~~~~-\frac{\partial}{\partial r}\left(\frac{a^4 + 2 r^4 + a^2 r(2 M + 3 r) + a^2\Delta \cos(2\theta)}{2\,\Delta\,\Sigma}\right)\frac{2\Delta^2 p_r^2 + 2\Delta\Sigma p_{\theta}^2}{\Sigma\left[a^4 + 2 r^4 + a^2 r(2 M + 3 r) + a^2\Delta cos2\theta\right]} \Biggr] \nonumber\\
		&&~~~~~~~- K_r (r - r_c)~,\label{4.4}\\
		&&\dot{\theta} = \frac{2 p_{\theta}\sqrt{\Delta}}{\Sigma\sqrt{p_{\theta}^2 + 2 p_r^2}}~,\label{4.5}\\
		&&\dot{p_{\theta}} = -\frac{1}{2E}\frac{\partial}{\partial \theta}\left(\frac{2\Delta^2 p_r^2 + 2\Delta\Sigma p_{\theta}^2}{\Sigma[a^4 + 2 r^4 + a^2 r(2 M + 3 r) + a^2 \Delta \cos2\theta]}\right) - K_{\theta}~r^2_{+}(\theta - \theta_c)~.	\label{4.6}
\end{eqnarray}
These are the equations which will be used to study the Lyapunov exponents in the later sections.

\section{Importance of analysing Lyapunov Exponent (LE)} \label{sec4}
The maximal Lyapunov exponent, which quantifies the average exponential rate of separation between nearby trajectories, is defined in general dynamical systems theory \cite{Sandri, Strogatz}. In our setup, this leads to the following standard expression, valid for almost all choices of initial separations
\begin{equation}
\lambda_{L,max}=\lim_{\tau\to\infty}\frac{1}{\tau}\ln\left(\frac{\delta r(\tau)}{\delta r(0)}\right)~,\label{LYP_max}
\end{equation}
where  $\delta r(\tau)$ denotes the separation between two initially close trajectories at time $\tau$, and $\delta r(0)$ denotes their initial separation. As we focus on null geodesics, $\tau$, the affine parameter, is implicitly defined through \ref{LYP_max}.

Let us note that in this classical setup, the maximal LE, $\lambda_L$ is bounded above by the surface gravity $\kappa$ of the black hole \cite{Hashimoto:2016dfz}. This bound is consistent with the temperature-based constraint introduced by Maldacena, Shenker, and Stanford (MSS) \cite{Maldacena_2016}, which reads
\begin{equation}
	\lambda_L \leq \frac{2\pi T}{\hbar}~,\label{MSS}
\end{equation}
where $\lambda_L$ denotes the maximal LE of the system, and $T$ is the associated temperature of the black hole.

Before proceeding with the numerical analysis of the LE, we first discuss the significance of its calculation in this context. The central goal is to probe the fundamental differences between extremal and nonextremal black holes, particularly in the context of whether extremal configurations can arise as continuous limits of near-extremal ones. While usual thermodynamic quantities such as temperature and entropy offer important insights, they often provide incomplete information when approaching extremality, especially due to the vanishing surface gravity and subtle near-horizon behavior in the extremal regime.

In this regard, dynamical indicators such as LEs serve as powerful tools to assess the intrinsic instability of particle trajectories in black hole spacetimes. As the maximal LE quantifies the rate at which two nearby trajectories diverge in phase space, and thus acts as a classical measure of chaos. Moreover, LE is related to the temperature of the system in the semi-classical limit and constrained by the MSS bound (\ref{MSS}), offering valuable insight into the system’s behavior as extremality is approached. 


\subsection{LE For RN black hole (for both nonextremal and extremal cases)}\label{subsec5.1}
We begin by analyzing the Lyapunov exponent (LE) for the motion of a massless particle in the Reissner–Nordström (RN) black hole background, focusing on how it varies with the black hole's charge. From the horizon structure of the RN black hole (\ref{RN_horizon}), it is evident that increasing the charge $Q$ leads to a decrease in the radius of the outer horizon $r_{+}$. We solve the associated dynamical equations of motion \Big[refer to \ref{3.4}, \ref{3.5}, \ref{3.6}, and \ref{3.7}\Big], employing the Runge-Kutta fourth-order scheme. The initial conditions for $r, p_{r}$ and $\theta$ are chosen as $3.5, 20.0$ and $0.0$ respectively, and the initial condition for $p_{\theta}$ is determined from the constant value of the system energy $E=50$. The other parameter values are chosen as $K_r=100$, $K_{\theta}=25$, $\theta_c=0$, and $r_c=4.3$ and mass of the RN black hole as $M=1.0$.

\begin{figure}[H]
	\centering
	\includegraphics[width=1.0\linewidth]{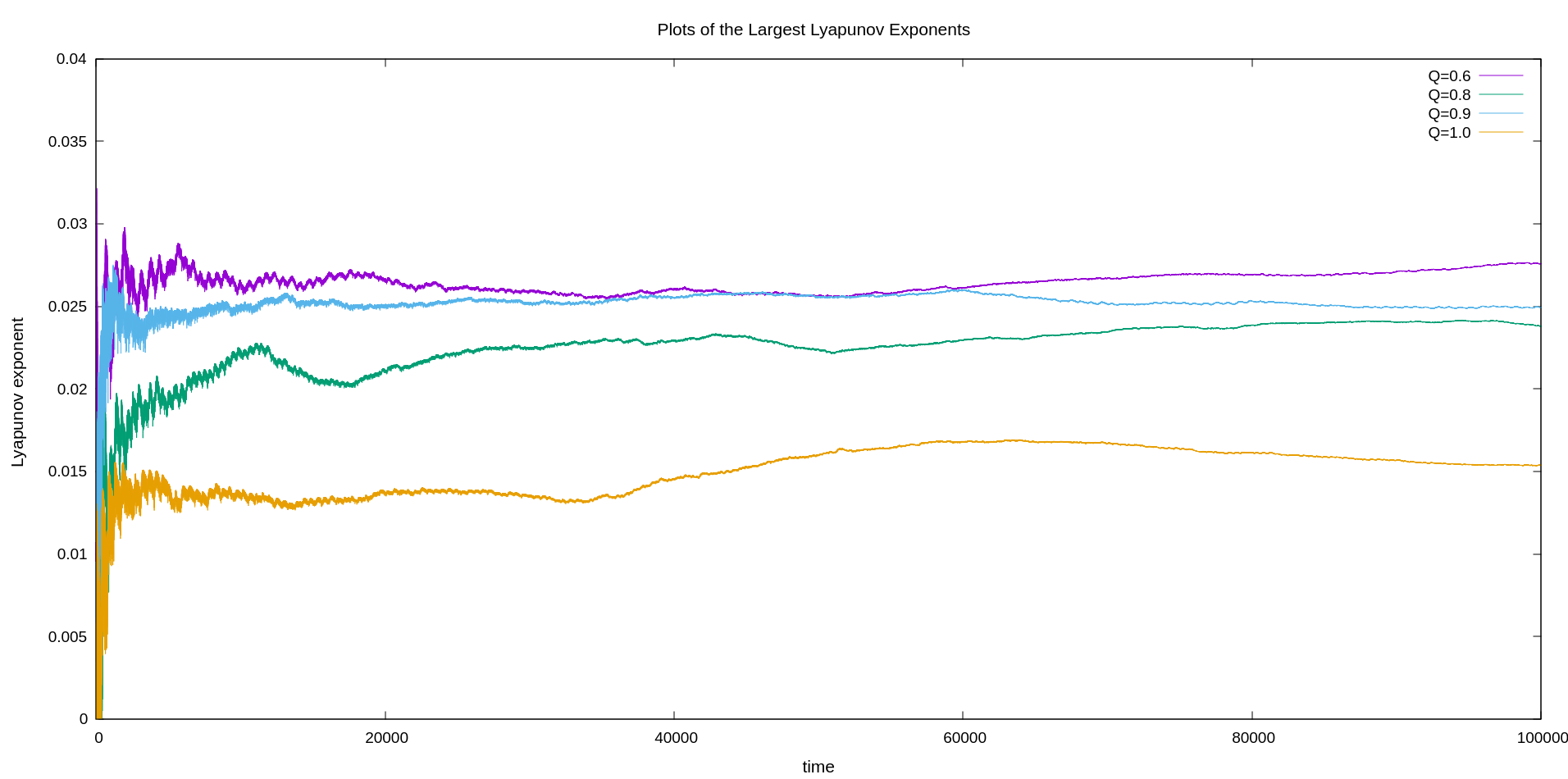}
	\caption{Plots of the largest Lyapunov exponent for Reissner-Nordström black holes with charges \( Q = 0.4, 0.5, 0.6, 0.8, 0.9 \), and the extremal case \( Q = 1.0 \). The exponent initially fluctuates but stabilizes to distinct saturation values, indicating reduced chaotic behavior as the extremal limit is approached.}
	\label{fig:1}
\end{figure}

\ref{fig:1} illustrates the behavior of the largest Lyapunov exponent for RN black holes with varying charges: $Q = 0.4, 0.5, 0.6, 0.8, 0.9$, as the system approaches the extremal limit at $Q = 1.0$. Initially, the Lyapunov exponents exhibit fluctuations but eventually settle to some saturation values. Notably, as the charge increases toward extremality, these saturation values gradually decrease (for $Q=0.6, 0.8$ and $ 0.9$, the largest LE values are respectively $0.027580, 0.023793$ and $0.024924$), ultimately reaching approximately $0.015380$ in the extremal case ($Q=M=1.0$). These values remain well below the well-established MSS bound, except in the extremal case where the bound is violated, with the Lyapunov exponent saturating at a positive value of approximately $\sim 0.015380$.

These results suggest that the motion of particles near an RN black hole becomes less chaotic as the black hole approaches the extremal limit. The drop in the Lyapunov exponent indicates that nearby particle trajectories diverge more slowly, pointing to a more stable dynamical system. In the extremal case, the chaos appears to be significantly reduced; nevertheless, a residual degree of chaos persists even though the surface gravity vanishes.


\begin{figure}[H]
	\centering
	\includegraphics[width=1.0\linewidth]{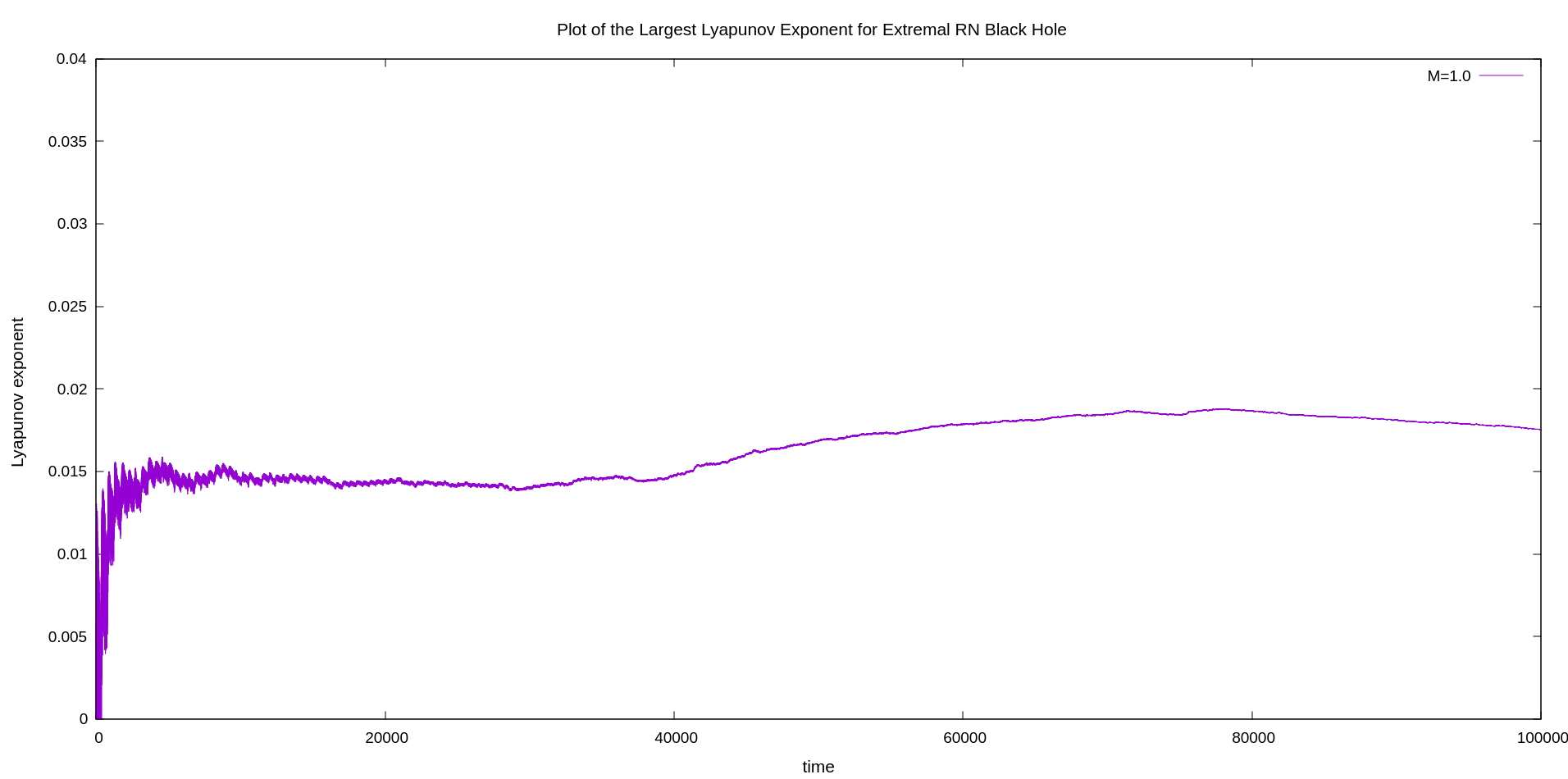}
    \caption{Largest Lyapunov exponent specifically for the extremal Reissner-Nordström black hole (\( Q = M = 1.0 \)). Significant initial fluctuations settle into a moderate saturation value (0.017518), demonstrating subtle yet persistent chaotic dynamics in the extremal configuration.}	
    \label{fig:2}
\end{figure}

In \ref{fig:2}, we specifically examine the extremal Reissner-Nordström black hole, characterized by the condition $Q = M = 1.0$ (see the metric in \ref{RN etremal metric}). The Lyapunov exponent in this scenario exhibits significant fluctuations at first, but eventually settles at a moderate saturation value of 0.017518. This value is close to, but slightly higher than, the saturation value obtained when approaching extremality from the non-extremal side. The near agreement between these two approaches indicates that the extremal configuration can be viewed as a smooth continuation of the non-extremal case, with only minor quantitative differences in the chaotic indicators.
Furthermore, in the extremal RN case, the Lyapunov exponent saturates at a finite positive value ($\lambda_{L}\sim 0.017519$), despite the vanishing surface gravity. This directly violates the MSS chaos bound, confirming that the extremal RN black hole follows a sustained residual chaotic dynamics beyond the expected thermodynamic limit.

Overall, these findings suggest that the motion of particles near RN black holes becomes progressively less chaotic as extremality is approached, and that the direct extremal background shares close dynamical characteristics with the near-extremal limit. Importantly, in all cases, the saturation values remain small compared to the MSS chaos bound, although the extremal case still demonstrates residual chaotic features despite the vanishing surface gravity. Therefore, it indicates one particular fact that extremal RN black holes are constituted with qualitatively distinct dynamical configurations from the non-extremal ones, as long as we are concerned about the chaos bound in the near-horizon regime.

\subsection{Analysis of Poincaré Sections for the RN System}

To further substantiate the presence of chaotic behavior in the Reissner–Nordström background and to complement the Lyapunov exponent analysis presented earlier, we now study the Poincaré sections of the system. The Poincaré map serves as a powerful diagnostic tool for nonlinear dynamical systems, allowing us to visualize the transition from regular to chaotic motion in the reduced phase space.

In our setup, a massless test particle moves in the RN spacetime under the influence of the total energy function given in \ref{3.3}, which includes harmonic confinement terms along both the radial and angular directions. The inclusion of these harmonic terms ensures bounded motion of the particle, allowing us to probe the near-horizon dynamics over long evolution times without the trajectory escaping to infinity. As discussed earlier, these confining potentials are static and non-dissipative, serving solely as a numerical device to stabilize the orbits and to map the intrinsic gravitationally induced instabilities near the horizon. The same approach has been widely employed in previous analyses of horizon-induced chaos \cite{Hashimoto:2016dfz,Dalui:2018,Dalui:2019,Bera_2022,Das_2024}.

We numerically integrate the coupled equations of motion [\ref{3.4}-\ref{3.7}] using a fourth-order Runge--Kutta method with a fixed step size $h = 0.01$. The parameters are set to $M = 1.0$, $K_{r} = 100$, $K_{\theta} = 25$, $r_{0} = 4.3$, and $\theta_{c} = 0$. The initial conditions for the variables are chosen as 
$r = 3.5$, $p_{r} = 20.0$, and $\theta = 0.0$, while $p_{\theta}$ is determined from the constant total energy $E = 50$. For each black hole charge $Q$, we solve the system for a large number of random initial conditions in a small neighborhood of the reference point to ensure a representative sampling of the accessible phase space.

The Poincaré map is constructed by recording $(r, p_{r})$ 
whenever the trajectory crosses $\theta = 0$ with $p_{\theta} > 0$ 
(single orientation). Regular motion produces smooth KAM tori, 
whereas chaotic motion yields distorted or broken tori and scattered points.

\begin{figure}[H]
  \centering
  \begin{subfigure}{0.48\textwidth}
    \includegraphics[width=\linewidth]{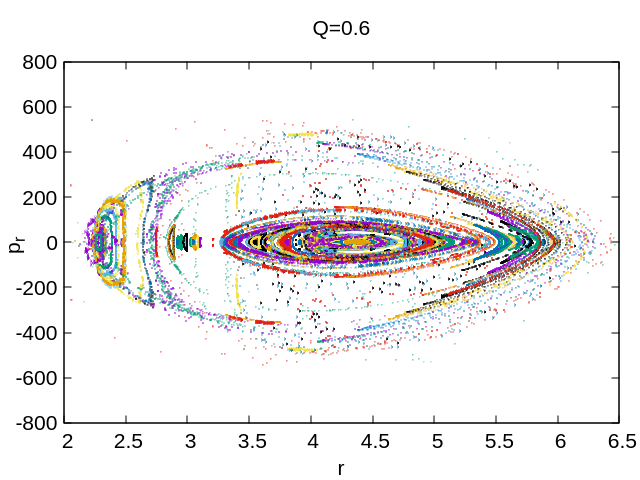}
    \caption{$Q = 0.6$}
  \end{subfigure}\hfill
  \begin{subfigure}{0.48\textwidth}
    \includegraphics[width=\linewidth]{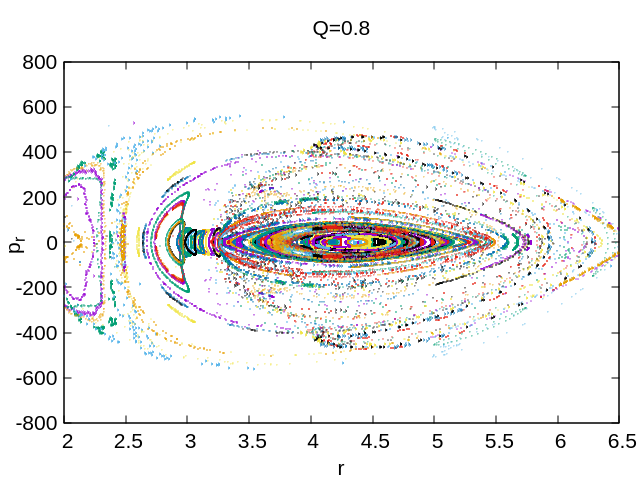}
    \caption{$Q = 0.8$}
  \end{subfigure}\\[0.8em]
  \begin{subfigure}{0.48\textwidth}
    \includegraphics[width=\linewidth]{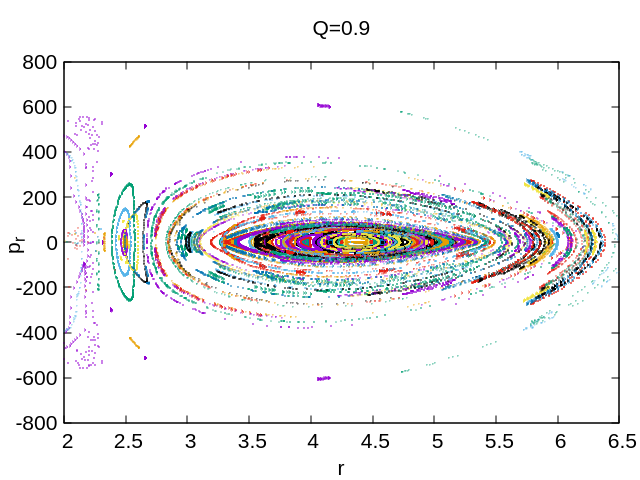}
    \caption{$Q = 0.9$}
  \end{subfigure}\hfill
  \begin{subfigure}{0.48\textwidth}
    \includegraphics[width=\linewidth]{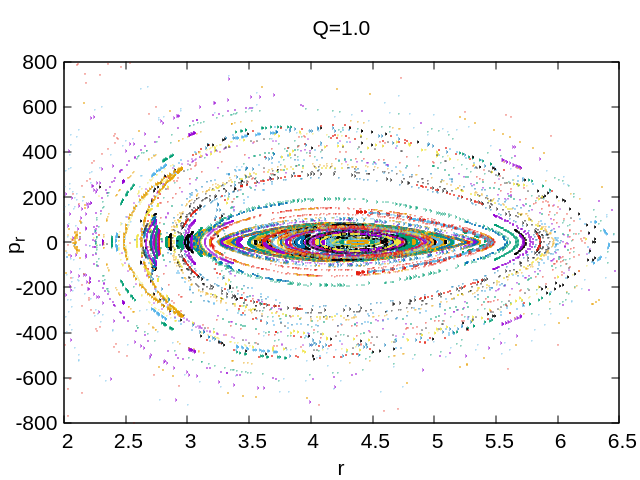}
    \caption{$Q = 1.0$ (extremal)}
  \end{subfigure}
  \caption{\textbf{Poincaré sections \((r, p_{r})\) for the Reissner--Nordström spacetime.}
  Orbits are recorded at $\theta = 0$ with $p_{\theta} > 0$.
  Parameters: $M = 1$, $K_{r} = 100$, $K_{\theta} = 25$, $E = 50$, $r_{c} = 4.3$, and $\theta_{c} = 0$.
  For a moderate charge ($Q = 0.6$), the section shows visible distortions and scattered points near the
  edges, indicating mild chaotic behavior. As $Q$ increases to $0.8$ and $0.9$, the KAM tori become
  progressively smoother and more regular. At the extremal value ($Q = 1.0$), the phase space is 
  dominated by well-formed nested tori, signifying the restoration of integrable motion. 
  This trend is consistent with the Lyapunov exponent analysis, where the degree of chaos 
  decreases as the charge increases and the horizon moves farther from the confined region.}
  \label{fig:RN_poincare}
\end{figure}

\ref{fig:RN_poincare} shows the sections for $Q = 0.6,\, 0.8,\, 0.9,\, 1.0$ where increasing $Q$ moves the relevant horizon farther from the harmonic oscillator center $r_{0}$. Consistently, the phase-space structures regularize with growing $Q$: for $Q = 0.6$ one observes noticeable torus distortions and scattered points near the outer lobes, while for $Q = 0.8$ and $Q = 0.9$ the broken tori progressively heal and the islands sharpen. At the extremal value $Q = 1.0$, the section is dominated by nested, smooth tori around $r_{0}$, with only sparse remnants of irregularity. This trend corroborates the Lyapunov-exponent behavior reported above: as the horizon recedes from the bounded integrable core, near-horizon nonlinearity weakens and the dynamics revert toward quasi-integrable motion.

\subsection{LE For Kerr black hole (for both nonextremal and extremal cases)}\label{subsec5.2}
The dynamical equations of motion \Big[refer to  \ref {4.3},  \ref {4.4}, \ref {4.5}, and \ref {4.6}\Big] are numerically solved employing the Runge-Kutta fourth-order scheme, keeping all the initial conditions and the parameter values the same as in the earlier section. 

\begin{figure}[H]
    \centering
    \includegraphics[width=1.0\linewidth]{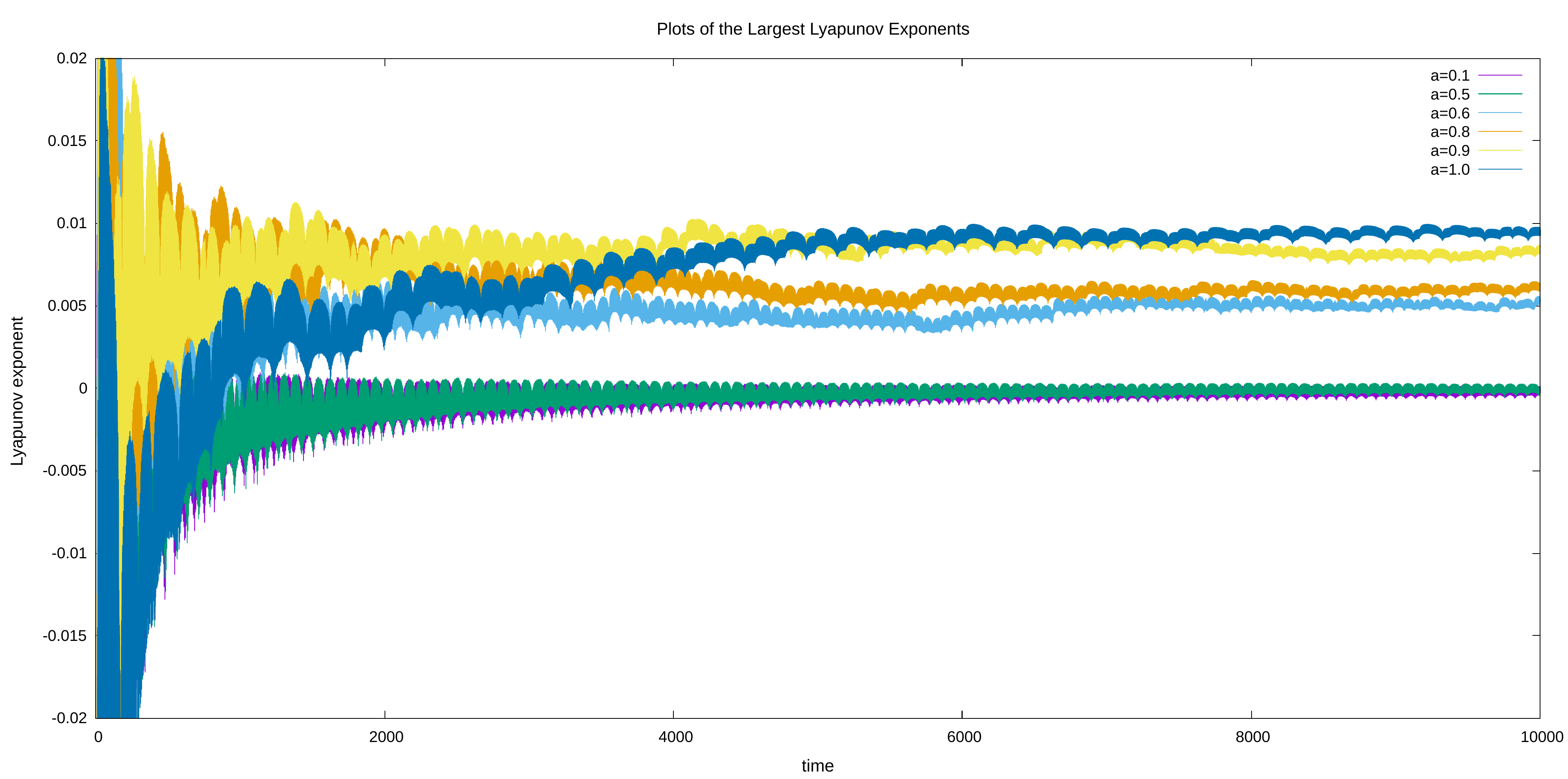}
    \caption{Evolution of the largest Lyapunov exponent for Kerr black holes with varying rotation parameters \( a = 0.1, 0.5, 0.6, 0.8, 0.9 \), and the extremal limit \( a = 1.0 \). The saturation values increase with spin, showing stronger chaotic behavior as the black hole approaches extremality.}    
    \label{fig:3}
\end{figure}

In \ref{fig:3}, we show how the largest Lyapunov exponent evolves over time for Kerr black holes (see the metric in \ref{kerr_metric}) with different rotation parameters: $a=0.1, 0.5, 0.6, 0.8, 0.9$, and the extremal limit $a=1.0$. For the nonextremal scenarios ($a<1.0$), the Lyapunov exponent initially fluctuates but then settles down to specific saturation values. As the rotation parameter increases and becomes closer to the extremal limit, these saturation values steadily increase (for $a=0.1, 0.5, 0.6, 0.8, 0.9$ we get the largest LE values respectively as $-0.0002, 0.0001, 0.0054, 0.0062, 0.0085$) which still respects the MSS bound but at the extremal limit ($a=M=1.0$) this value saturates at 0.0095 which clearly violates the MSS bound in this scenario. This clearly means that chaotic behavior gets stronger as the black hole spins faster. This trend suggests that rotation acts as a destabilizing factor in the dynamics of test particles near the Kerr black hole. Moreover, the monotonic growth of the Lyapunov exponent with increasing spin indicates a direct link between rotational energy and the system’s sensitivity to initial conditions.
These findings reinforce the idea that rotation amplifies dynamical instability in black hole spacetimes. Unlike the RN case, where chaos diminishes as extremality is approached, the Kerr scenario demonstrates that extremality, achieved through angular momentum, enhances chaos. This distinction highlights the fundamentally different roles played by electric charge and angular momentum in governing the dynamics of spacetime.
\begin{figure}[H]
    \centering
    \includegraphics[width=1.0\linewidth]{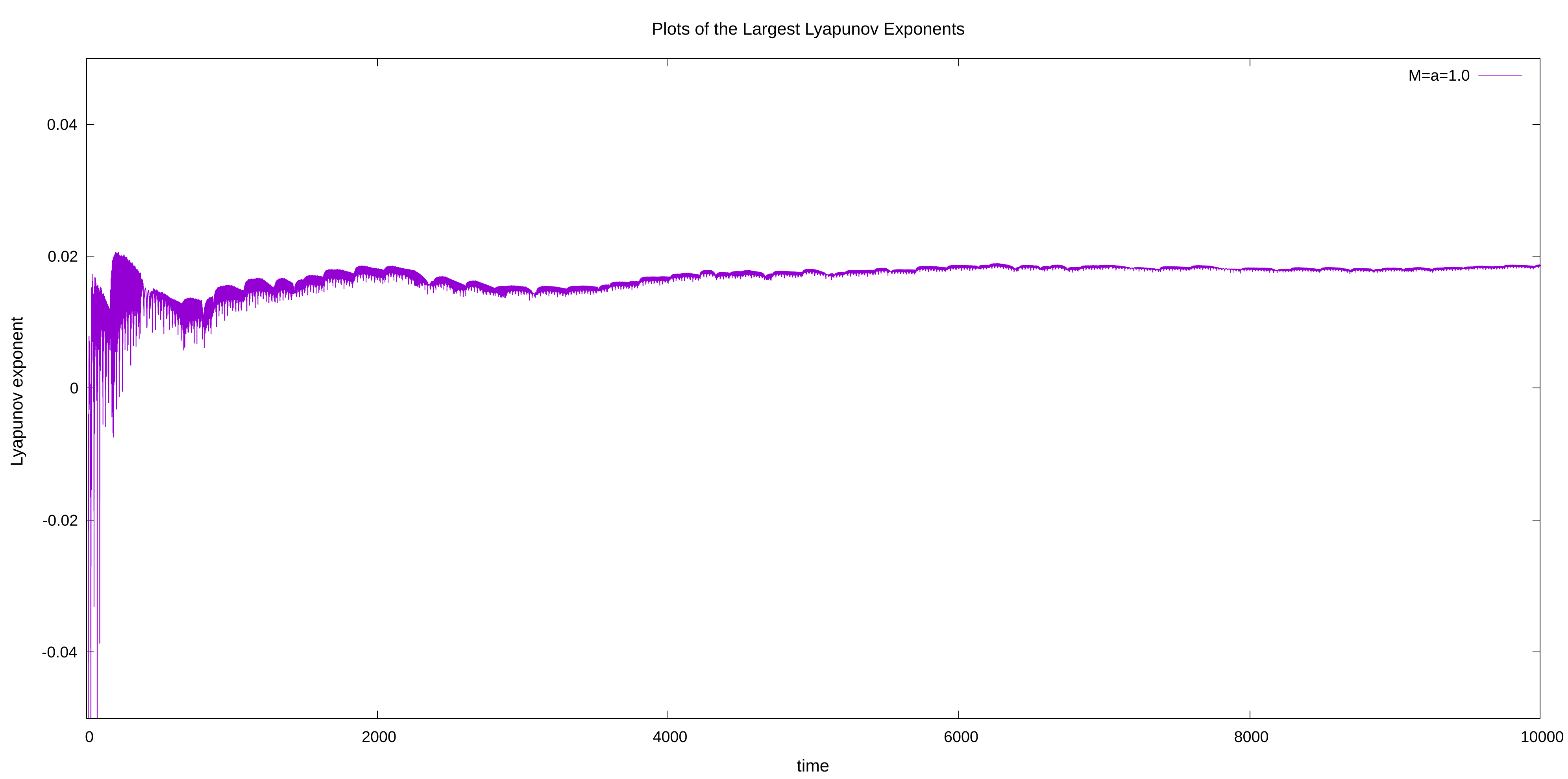}
    \caption{Lyapunov exponent for the extremal Kerr black hole (\( a = M = 1.0 \)). Noticeable fluctuations initially appear before stabilizing at a significantly higher saturation value (0.0185), highlighting enhanced chaotic and unstable dynamics in the fastest spinning Kerr black hole scenario.}    
    \label{fig:4}
\end{figure}
In \ref{fig:4}, we specifically examine the extremal Kerr black hole case (see the metric in \ref{kerr_extremal}), characterized by the condition $a = M = 1.0$. After significant initial fluctuations, the Lyapunov exponent settles to a saturation value of 0.0185. While this value is somewhat higher than the near-extremal case, the difference is moderate. Importantly, both results remain within the same order of magnitude, suggesting that the extremal configuration shares close dynamical properties with the limiting non-extremal case. However, one thing to notice here is that, in this case, the extremal Kerr black hole exhibits a finite positive Lyapunov exponent ($\lambda_{L}\sim 0.0185$), even at zero Hawking temperature, just like we have seen in the case of the extremal RN black hole in the previous section. This constitutes another clear violation of the MSS chaos bound, showing that extremal Kerr black holes support an even stronger chaotic regime than their non-extremal counterparts. Similar violations of the chaos bound have been observed in other contexts, such as for charged particle near Kerr–Newman–AdS black holes \cite{Gwak:2022xje} and for homoclinic particle orbits in black holes with anisotropic matter fields \cite{Jeong:2023hom}. Therefore, our analysis reveals that extremal geometry represents another universal setting in which such violations naturally occur.

Thus, unlike the RN scenario where chaos diminishes toward extremality, the Kerr case demonstrates a mild strengthening of chaotic features with increasing spin. The comparison between the near-extremal and exact extremal situations indicates continuity in behavior, with only quantitative differences in the degree of chaos. This suggests that the extremal Kerr geometry can be reasonably interpreted as a smooth extension of the non-extremal family, but with enhanced chaotic instability at exact extremality. Importantly, for the extremal Kerr black hole the Lyapunov exponent saturates at a finite positive value despite the vanishing Hawking temperature. This clear violation of the MSS chaos bound reveals that extremal black holes, whether it is a Kerr or an  RN black hole, harbour a totally different dynamical structure in their near-horizon region from the maximal chaos bound perspective with respect to the non-extremal ones.


\subsection{Analysis of Poincaré Sections for Kerr system}

To complement the Lyapunov exponent analysis and visualize the emergence of chaos in the Kerr spacetime, we now present the corresponding Poincaré sections. The same numerical setup discussed earlier is used here. A massless test particle evolves according to the total energy function \ref{4.2}, 
which includes the static harmonic confinements along the $r$ and $\theta$ directions. These potentials ensure that the motion remains bounded, allowing for the identification of chaotic signatures in the near-horizon region over extended integration times. 

The equations of motion [\ref{4.3}-\ref{4.6}] are solved using a fourth-order Runge-Kutta algorithm with a fixed step size $h = 0.01$. Unless otherwise stated, the parameters are set as 
$M = 1.0$, $K_{r} = 100$, $K_{\theta} = 25$, 
$r_{c} = 4.3$, $\theta_{c} = 0$, and the total energy $E = 50$. 
Initial conditions are chosen as $r = 3.5$, $p_{r} = 20.0$, and $\theta = 0.0$, while $p_{\theta}$ is obtained from the conserved energy constraint. The Poincaré map is constructed by recording $(r, p_{r})$ each time the trajectory crosses $\theta = 0$ with $p_{\theta} > 0$.

\begin{figure}[H]
  \centering
  \begin{subfigure}{0.32\textwidth}
    \includegraphics[width=\linewidth]{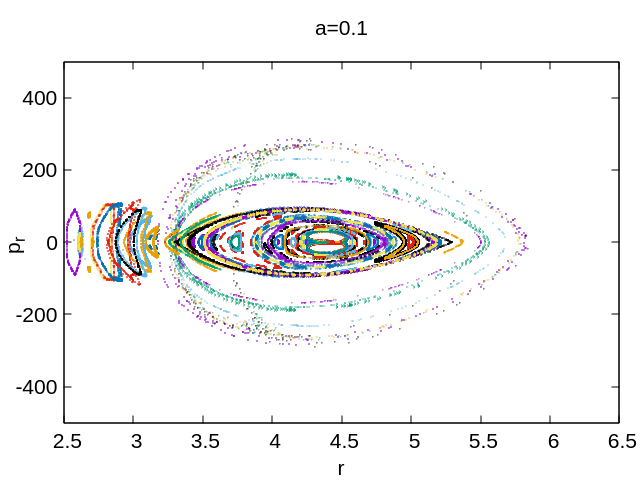}
    \caption{$a = 0.1$}
  \end{subfigure}\hfill
  \begin{subfigure}{0.32\textwidth}
    \includegraphics[width=\linewidth]{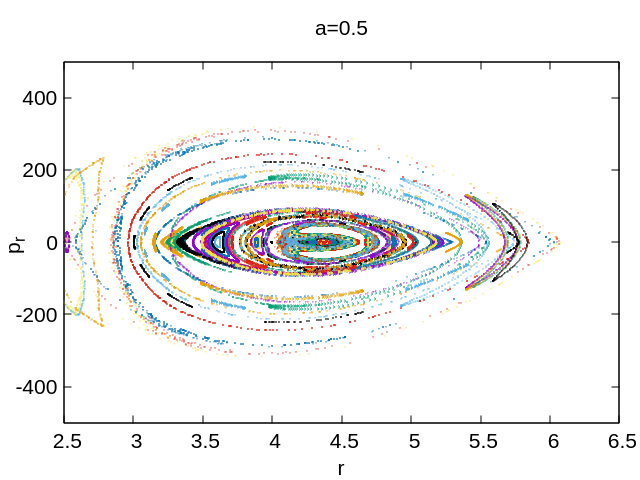}
    \caption{$a = 0.5$}
  \end{subfigure}\hfill
  \begin{subfigure}{0.32\textwidth}
    \includegraphics[width=\linewidth]{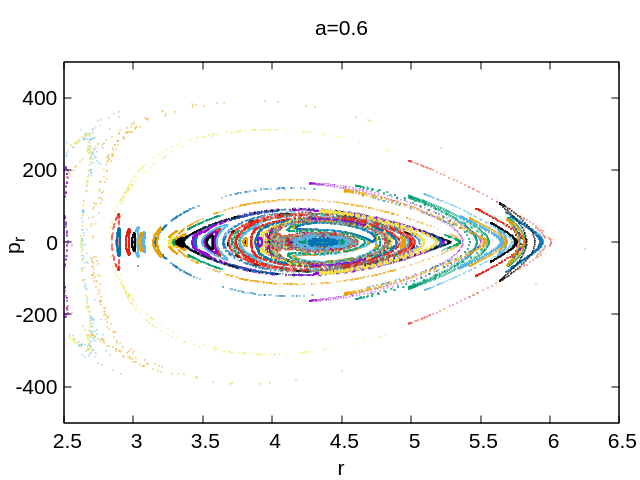}
    \caption{$a = 0.6$}
  \end{subfigure}\\[0.6em]
  \begin{subfigure}{0.32\textwidth}
    \includegraphics[width=\linewidth]{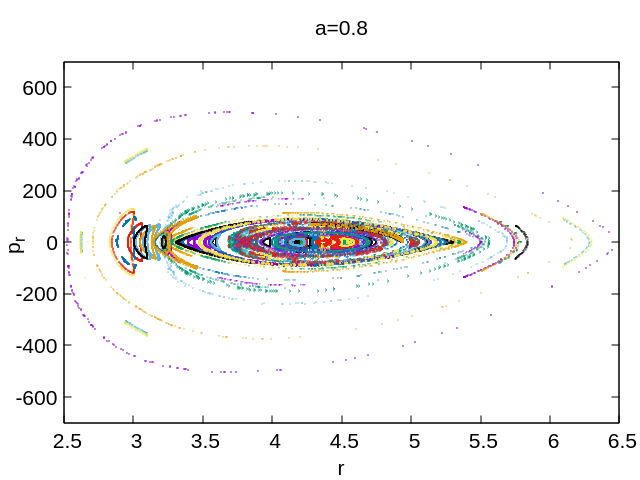}
    \caption{$a = 0.8$}
  \end{subfigure}\hfill
  \begin{subfigure}{0.32\textwidth}
    \includegraphics[width=\linewidth]{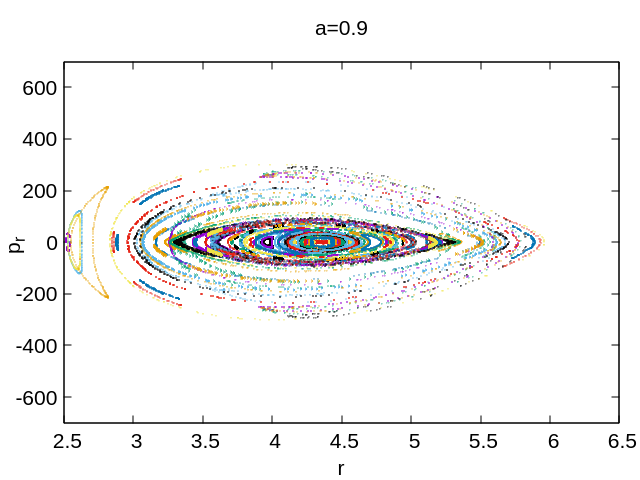}
    \caption{$a = 0.9$}
  \end{subfigure}\hfill
  \begin{subfigure}{0.32\textwidth}
    \includegraphics[width=\linewidth]{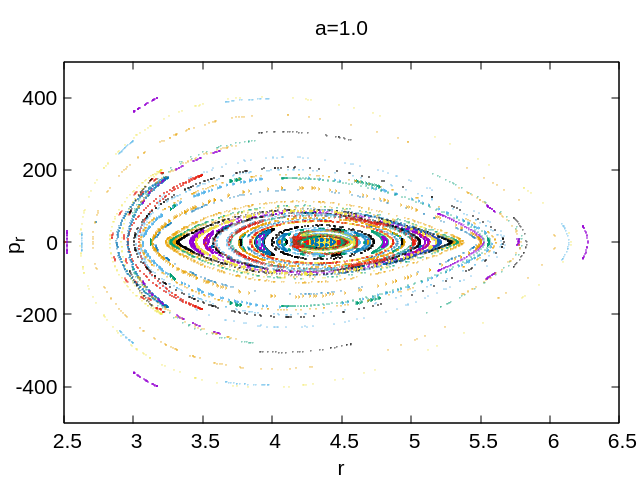}
    \caption{$a = 1.0$ (extremal)}
  \end{subfigure}
  \caption{\textbf{Poincaré sections \((r,p_{r})\) for the Kerr spacetime.}
  Orbits are recorded at $\theta = 0$ with $p_{\theta} > 0$. 
  Parameters: $M = 1$, $K_{r} = 100$, $K_{\theta} = 25$, 
  $E = 50$, $r_{c} = 4.3$, and $\theta_{c} = 0$.
  At low spin ($a=0.1$), the section shows regular nested KAM tori.
  With increasing $a$, the tori become progressively distorted and fragmented, 
  and at the extremal limit ($a = 1.0$), most KAM structures are destroyed, 
  leaving a predominantly chaotic phase space consistent with the positive 
  Lyapunov exponents observed in Sec.~4.2.}
  \label{fig:Kerr_poincare}
\end{figure}

 \ref{fig:Kerr_poincare} displays the resulting Poincaré sections for different values of the rotation parameter $a = 0.1, 0.5, 0.6, 0.8, 0.9,$ and $1.0$. At small spin ($a = 0.1$), the phase space is dominated by well-formed KAM tori, indicating nearly integrable motion. As $a$ increases to $0.5$ and $0.6$, the tori begin to distort, and isolated islands of instability appear near the outer regions of phase space. When the spin increases further to $a = 0.8$ and $0.9$, the tori become progressively broken, and dense scattered points emerge, signaling the onset of chaotic dynamics. In the extremal case ($a = 1.0$), the phase space becomes largely irregular, with most KAM structures destroyed and only small regular islands surviving near the center. This trend is consistent with the behavior of the Lyapunov exponents presented earlier, where higher spin enhances the system’s dynamical instability and 
amplifies the near-horizon nonlinear effects.

These results establish a clear correspondence between rotation and chaos: while the RN system becomes more regular as the horizon moves away from the bounded region, the Kerr system exhibits the opposite tendency, i.e. increasing spin strengthens the coupling between angular and radial motions, thereby promoting chaotic evolution.

\noindent
\textbf{Remark on the role of the harmonic confinement:} 
As emphasized throughout this work, the harmonic potential is introduced purely as a numerical tool to confine trajectories, allowing horizon-induced instabilities to manifest through bounded recurrent motion. To further clarify that this confinement does not \emph{create} chaos but only makes it observable in a bounded phase space, we also computed the Lyapunov exponents for both the Reissner--Nordström and Kerr 
black hole systems \emph{without} including the harmonic potential.  The results, shown in Appendix~\ref{app:withoutHO}, reveal that the largest 
Lyapunov exponents remain positive, following the same parametric trend with charge  ($Q$) and spin ($a$) as discussed in the main text, although their absolute values are significantly smaller. This confirms that the exponential sensitivity of nearby geodesics is an intrinsic property of the near-horizon geometry, while the harmonic trap merely provides a bounded setting to visualize its full chaotic manifestation.\\

\noindent
\textbf{Astrophysical motivation:}
In realistic astrophysical environments, black holes are rarely isolated systems. They may be surrounded by accretion disks, dark matter halos, magnetic fields, or experience external tidal interactions due to nearby compact objects. Such environmental effects can introduce additional forces that effectively confine particle motion or induce restoring behavior around equilibrium
configurations. From this perspective, the harmonic confinement adopted in \ref{3.3} can be viewed as a minimal phenomenological representation of these external interactions. This allows us to investigate how the intrinsic near-horizon instabilities of black hole spacetimes manifest when particle motion becomes bounded, without committing to a specific astrophysical model.

\section{Conclusion} \label{conclusion}
Over the years, the study of black holes has attracted increasing attention within the research community. In particular, black hole thermodynamics has emerged as a powerful framework, providing alternative perspectives on the nature of gravitation. Extremal black holes, as special cases characterized by extremal charge or spin, serve as critical testing grounds due to their distinctive properties. Notably, they exhibit vanishing Hawking temperature while maintaining finite entropy, in agreement with the statistical mechanics interpretation.

It is important to emphasize that the vanishing temperature of extremal black holes typically arises when taking the extremal limit of results originally derived for nonextremal cases. This raises a deeper question: are extremal black holes merely limiting cases of nonextremal solutions, or do they constitute fundamentally distinct entities? This issue remains a topic of active debate in the literature.

 In this work, we have undertaken a detailed investigation of the near-horizon dynamics of a massless particle in both rotating and non-rotating black hole spacetimes, with a particular emphasis on identifying and characterizing signatures of classical chaos. Our study was motivated by the observation that extremal black holes occupy a particularly singular and intriguing position in the field of gravitational physics. Unlike their nonextremal counterparts, extremal black holes are defined by vanishing surface gravity and thus zero temperature, accompanied by an emergent near-horizon throat geometry that is often described in terms of an $AdS_2 \times S^2$ structure in four dimensions or its generalizations. These features give rise to unique thermodynamic and geometric properties, yet the dynamical behavior of particles and fields in such backgrounds remains relatively less understood. Understanding the extent to which chaotic motion persists or is modified in these spacetimes is therefore a key step toward bridging classical dynamics, black hole thermodynamics, and microscopic descriptions of black hole states.

To address this question, we pursued two complementary and independent routes to extremality. In the first approach, we began with generic nonextremal geometries and systematically approached the extremal limit, allowing us to capture the dynamical features that emerge as the horizon temperature tends to zero. In the second approach, we analyzed the extremal metric directly, without appealing to any limiting procedure, thereby providing a conceptually distinct framework for probing particle dynamics in extremal geometries. Obtaining the same results from both approaches makes our conclusions more reliable and helps us distinguish the true features of extreme black holes from effects that arise solely from taking the limit.

Our analysis shows that the Lyapunov exponents, which measure the instability and chaos in dynamical systems, exhibit qualitatively similar patterns in both approaches. In particular, we find that the growth rates of perturbations do not diverge in the extremal limit but instead remain finite and saturate at well-defined values. This behavior is highly nontrivial, as one might have anticipated either a suppression of chaotic growth due to the vanishing surface gravity or, alternatively, a divergent response associated with the infinite redshift near extremal horizons. Instead, the growth rates stabilize in a controlled fashion, leading to a universal structure that is markedly distinct from the behavior of nonextremal black holes. We have found that, the Lyapunov exponent remains finite and positive even at zero temperature contrary to a naive extrapolation of the Maldacena-Shenker-Stanford (MSS) chaos bound, For Reissner-Nordstr\"{o}m black holes, chaotic behavior diminishes but persists at extremality. In contrast, for Kerr black holes, it becomes stronger with increasing spin. These findings reveal that extremal black holes exhibit residual chaotic dynamics that effectively evade the MSS bound, distinguishing them as qualitatively different dynamical phases of gravity. Our results are consistent with and complementary to previous works that identified violations of the MSS chaos bound for the extremal scenarios in diverse black hole settings. In particular, studies of charged particles near Kerr–Newman–AdS spacetimes have shown that rotation and extremality promote bound violation \cite{Gwak:2022xje}, while analyses of homoclinic orbits in anisotropic black hole geometries have demonstrated similar effects arising from matter fields \cite{Jeong:2023hom}. Taken together, these works and our results emphasize that extremal horizons constitute a universal setting where the MSS bound is generically violated, irrespective of asymptotics or matter content. Therefore, our results underscore the need for a refined theoretical framework to comprehend quantum chaos in zero-temperature holographic systems and to elucidate the role of extremal horizons within the broader context of gravitational dynamics.

We emphasize that the chaotic behavior analyzed in this work arises entirely from classical particle dynamics in curved spacetime. The Lyapunov exponents computed here quantify classical sensitivity to initial conditions and should be distinguished
from quantum measures of chaos, such as those defined through out-of-time-order correlators (OTOC). In contrast, the MSS bound on chaos is a quantum bound associated with thermal many-body systems and information scrambling, with black hole temperature playing a central role.

The comparison made in this work should therefore be interpreted at a heuristic level. Near black hole horizons, classical trajectories exhibit exponential instability governed by the surface gravity, while the same geometric quantity sets the Hawking temperature that controls quantum scrambling rates. A similar situation arises in black hole thermodynamics, where quantities such as temperature and entropy were originally motivated through formal analogies with the laws of thermodynamics before a microscopic quantum interpretation was established. This shared geometric origin
motivates a qualitative comparison between classical Lyapunov exponents and the MSS bound, without implying a direct equivalence between classical and quantum chaos. Establishing a precise correspondence between these two notions remains an open problem and lies beyond the scope of the present analysis.

Finally, we note that in realistic astrophysical environments, black holes may coexist with surrounding matter fields, leading to effective geometries beyond the vacuum Reissner-Nordström and Kerr solutions. In such scenarios, additional interactions can naturally induce bounded particle motion and may enhance or modify the chaotic signatures discussed in this work. Our analysis may therefore find broader applicability in black holes dressed by anisotropic matter fields or external sources. Recent studies on rotating and charged rotating black holes with anisotropic matter
distributions~\cite{Kim:2019hfp,Kim:2021vlk,Kim:2025sdj} provide concrete examples of such backgrounds where these effects could become relevant.

\section*{Acknowledgements}
The research of KB is supported by the New Faculty Seed Grant (NFSG) of BITS Pilani Dubai Campus.

\appendix
\section{Lyapunov Exponents Without Harmonic Confinement}
\label{app:withoutHO}

To verify that the observed chaotic signatures do not arise solely from the inclusion 
of the harmonic potential, we recalculated the largest Lyapunov exponents (LEs) for 
both the Reissner-Nordström and Kerr black holes in the absence of any confining term. 
The numerical procedure follows the same integration scheme described in the main text, 
but with the external potential terms set to zero. For better visibility, the computed 
LE values have been rescaled by multiplying each data set by $10^{5}$. 

\begin{figure}[H]
    \centering
    \includegraphics[width=0.9\textwidth]{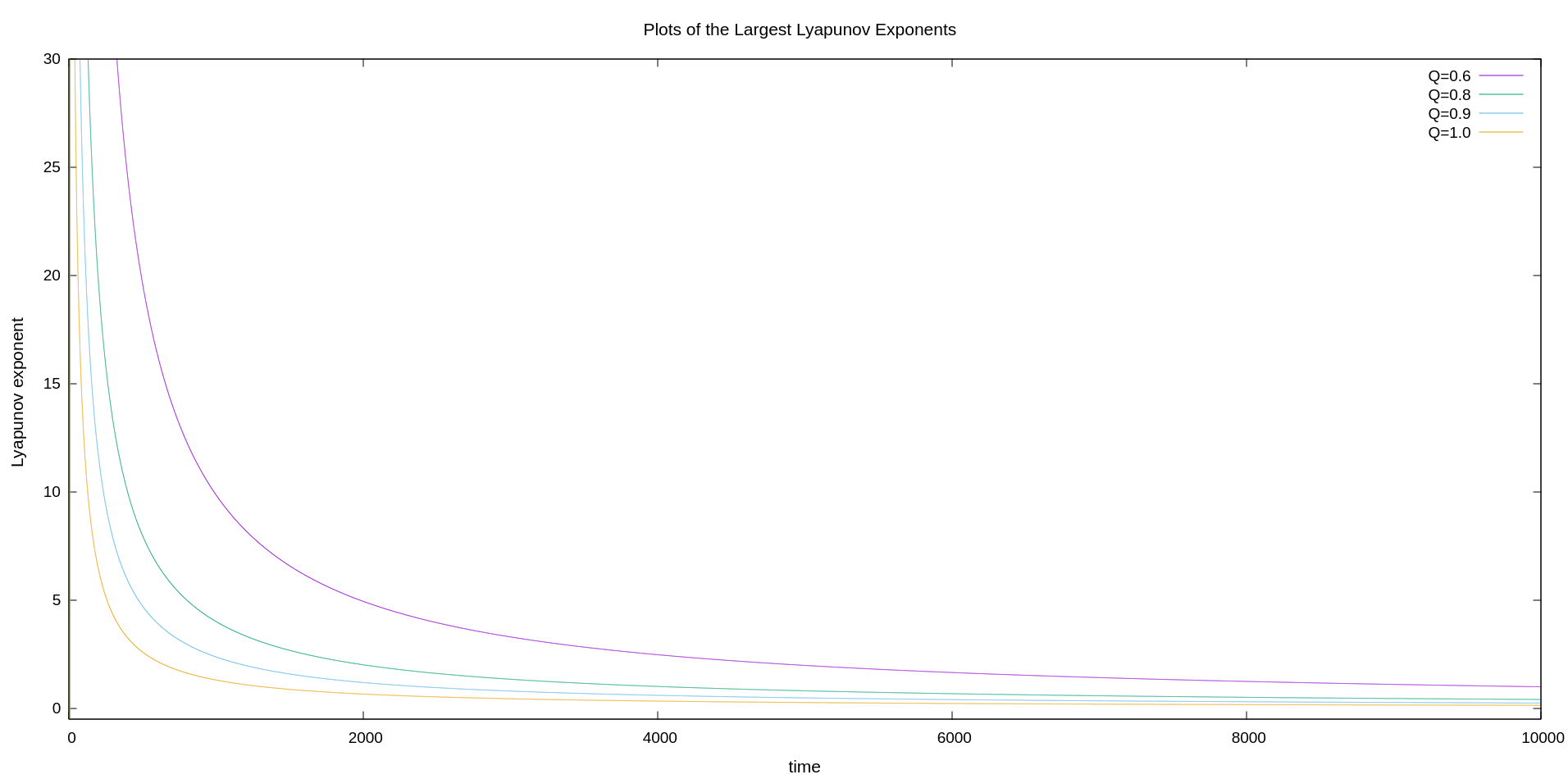}
    \caption{\textbf{Largest Lyapunov exponents for the Reissner--Nordström system 
    without harmonic confinement.}
    The data are rescaled by a factor of $10^{5}$ for clarity. 
    Despite smaller magnitudes, the exponents remain positive and decrease with increasing 
    charge $Q$, indicating that the exponential divergence of nearby geodesics persists 
    even without confinement. This demonstrates that the instability is intrinsic to the 
    near-horizon geometry rather than an artifact of the external potential.}
    \label{fig:LE_RN_noHO}
\end{figure}

\begin{figure}[H]
    \centering
    \includegraphics[width=0.9\textwidth]{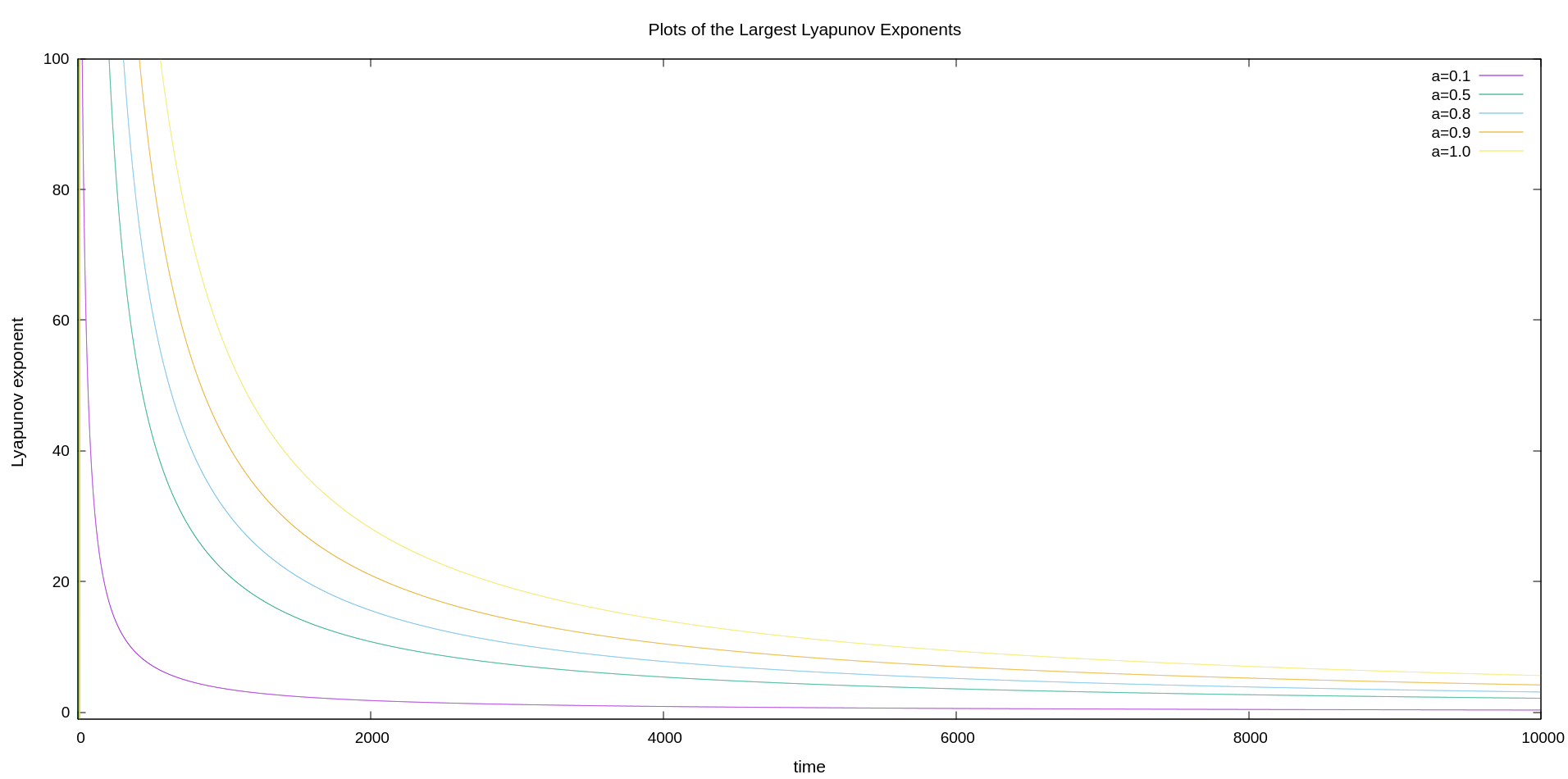}
    \caption{\textbf{Largest Lyapunov exponents for the Kerr system without harmonic confinement.}
    Each curve is rescaled by a factor of $10^{5}$. 
    The exponents grow with increasing spin parameter $a$, consistent with the trend observed 
    in the harmonically confined case. Although the magnitudes are smaller, the persistence of 
    positive values confirms that the rotationally enhanced instability is intrinsic to the 
    Kerr geometry itself.}
    \label{fig:LE_Kerr_noHO}
\end{figure}

\ref{fig:LE_RN_noHO} presents the variation of the largest Lyapunov exponents 
for the RN system for different charge values $Q = 0.6, 0.8, 0.9$, and $1.0$. 
Although the absolute magnitudes of the exponents are small, they remain positive and 
display the same decreasing trend with increasing $Q$ as in the harmonically confined case. 
Similarly, \ref{fig:LE_Kerr_noHO} shows the corresponding results for the Kerr system 
for several spin parameters $a = 0.1, 0.5, 0.8, 0.9,$ and $1.0$. Here, the LEs increase 
monotonically with $a$, again following the same qualitative behavior as reported in the main text. 
These results confirm that the underlying exponential instability originates from the near-horizon 
geometry itself, and that the harmonic confinement only enhances and bounds this behavior for 
clearer numerical diagnosis.

\bibliography{mybibliography_whm.bib}

@article{Dalui_2020,
   title={Horizon induces instability locally and creates quantum thermality},
   volume={102},
   ISSN={2470-0029},
   url={http://dx.doi.org/10.1103/PhysRevD.102.044006},
   DOI={10.1103/physrevd.102.044006},
   number={4},
   journal={Physical Review D},
   publisher={American Physical Society (APS)},
   author={Dalui, Surojit and Majhi, Bibhas Ranjan and Mishra, Pankaj},
   year={2020},
   month=aug }

@article{Dalui_2020_1,
   title={Near-horizon local instability and quantum thermality},
   volume={102},
   ISSN={2470-0029},
   url={http://dx.doi.org/10.1103/PhysRevD.102.124047},
   DOI={10.1103/physrevd.102.124047},
   number={12},
   journal={Physical Review D},
   publisher={American Physical Society (APS)},
   author={Dalui, Surojit and Majhi, Bibhas Ranjan},
   year={2020},
   month=dec }

@article{Dalui_2022,
   title={Horizon thermalization of Kerr black hole through local instability},
   volume={826},
   ISSN={0370-2693},
   url={http://dx.doi.org/10.1016/j.physletb.2022.136899},
   DOI={10.1016/j.physletb.2022.136899},
   journal={Physics Letters B},
   publisher={Elsevier BV},
   author={Dalui, Surojit and Majhi, Bibhas Ranjan},
   year={2022},
   month=mar, pages={136899} }

@article{Maldacena_2016,
   title={A bound on chaos},
   volume={2016},
   ISSN={1029-8479},
   url={http://dx.doi.org/10.1007/JHEP08(2016)106},
   DOI={10.1007/jhep08(2016)106},
   number={8},
   journal={Journal of High Energy Physics},
   publisher={Springer Science and Business Media LLC},
   author={Maldacena, Juan and Shenker, Stephen H. and Stanford, Douglas},
   year={2016},
   month=aug }

@book{Strogatz,
   title =     {Nonlinear dynamics and Chaos: with applications to physics, biology, chemistry, and engineering},
   author =    {Steven H. Strogatz},
   publisher = {Addison-Wesley Pub},
   isbn =      {0201543443,9780201543445,9780738204536,0738204536},
   year =      {1994},
   series =    {Studies in nonlinearity},
   edition =   {},
   volume =    {},
   url =       {http://gen.lib.rus.ec/book/index.php?md5=fa4190d4e05f55134df61c889ad304b0}
}

@article{Sandri,
author = {Sandri, Marco},
year = {1996},
month = {01},
pages = {},
title = {Numerical calculation of Lyapunov exponents},
volume = {6},
journal = {Math. J.}
}

@article{Das_2024,
   title={Near-horizon chaos beyond Einstein gravity},
   volume={110},
   ISSN={2470-0029},
   url={http://dx.doi.org/10.1103/PhysRevD.110.124037},
   DOI={10.1103/physrevd.110.124037},
   number={12},
   journal={Physical Review D},
   publisher={American Physical Society (APS)},
   author={Das, Surajit and Dalui, Surojit and Samanta, Rickmoy},
   year={2024},
   month=dec }

@article{Bera_2022,
   title={Quantum corrections enhance chaos: Study of particle motion near a generalized Schwarzschild black hole},
   volume={829},
   ISSN={0370-2693},
   url={http://dx.doi.org/10.1016/j.physletb.2022.137033},
   DOI={10.1016/j.physletb.2022.137033},
   journal={Physics Letters B},
   publisher={Elsevier BV},
   author={Bera, Avijit and Dalui, Surojit and Ghosh, Subir and Vagenas, Elias C.},
   year={2022},
   month=jun, pages={137033} }

@misc{reissner_h_1916_1447315,
  author       = {Reissner, H.},
  title        = {{Über die Eigengravitation des elektrischen Feldes 
                   nach der Einsteinschen Theorie}},
  month        = jan,
  year         = 1916,
  publisher    = {Zenodo},
  doi          = {10.1002/andp.19163550905},
  url          = {https://doi.org/10.1002/andp.19163550905}
}

@book{book:15209,
   title =     {The mathematical theory of black holes},
   author =    {S. Chandrasekhar},
   publisher = {Clarendon Press; Oxford University Press},
   isbn =      {0198512910,9780198512912},
   year =      {1983},
   series =    {The International series of monographs on physics 69},
   edition =   {1ST},
   volume =    {},
   url =       {http://gen.lib.rus.ec/book/index.php?md5=501485eb390043e51dc7a1bd85c9de7b}
}

@article{Gibbons:1994ff,
    author = "Gibbons, G. W. and Kallosh, R. E.",
    title = "{Topology, entropy and Witten index of dilaton black holes}",
    eprint = "hep-th/9407118",
    archivePrefix = "arXiv",
    reportNumber = "NI-94003, SU-ITP-94-9",
    doi = "10.1103/PhysRevD.51.2839",
    journal = "Phys. Rev. D",
    volume = "51",
    pages = "2839--2862",
    year = "1995"
}

@article{Hawking:1994ii,
    author = "Hawking, S. W. and Horowitz, Gary T. and Ross, Simon F.",
    title = "{Entropy, Area, and black hole pairs}",
    eprint = "gr-qc/9409013",
    archivePrefix = "arXiv",
    reportNumber = "NI-94-012, DAMTP-R-94-26, UCSBTH-94-25",
    doi = "10.1103/PhysRevD.51.4302",
    journal = "Phys. Rev. D",
    volume = "51",
    pages = "4302--4314",
    year = "1995"
}

@article{Teitelboim:1994az,
    author = "Teitelboim, Claudio",
    title = "{Action and entropy of extreme and nonextreme black holes}",
    eprint = "hep-th/9410103",
    archivePrefix = "arXiv",
    reportNumber = "IASSNS-HEP-94-84",
    doi = "10.1103/PhysRevD.52.6201",
    journal = "Phys. Rev. D",
    volume = "51",
    pages = "4315",
    year = "1995",
    note = "[Erratum: Phys.Rev.D 52, 6201 (1995)]"
}

@article{Sen:2008vm,
    author = "Sen, Ashoke",
    title = "{Quantum Entropy Function from AdS(2)/CFT(1) Correspondence}",
    eprint = "0809.3304",
    archivePrefix = "arXiv",
    primaryClass = "hep-th",
    doi = "10.1142/S0217751X09045893",
    journal = "Int. J. Mod. Phys. A",
    volume = "24",
    pages = "4225--4244",
    year = "2009"
}

@article{Strominger:1996sh,
    author = "Strominger, Andrew and Vafa, Cumrun",
    title = "{Microscopic origin of the Bekenstein-Hawking entropy}",
    eprint = "hep-th/9601029",
    archivePrefix = "arXiv",
    reportNumber = "HUTP-96-A002, RU-96-01",
    doi = "10.1016/0370-2693(96)00345-0",
    journal = "Phys. Lett. B",
    volume = "379",
    pages = "99--104",
    year = "1996"
}

@article{Nanda:2022szu,
    author = "Nanda, Pritam and Singha, Chiranjeeb and Tripathy, Pabitra and Ghosh, Amit",
    title = "{Hawking radiation as quantum mechanical reflection}",
    eprint = "2203.06588",
    archivePrefix = "arXiv",
    primaryClass = "gr-qc",
    doi = "10.1007/s10714-022-03007-1",
    journal = "Gen. Rel. Grav.",
    volume = "54",
    number = "10",
    pages = "120",
    year = "2022"
}

@article{Ghosh:1995rv,
    author = "Ghosh, Amit and Mitra, P.",
    title = "{Temperatures of extremal black holes}",
    eprint = "gr-qc/9507032",
    archivePrefix = "arXiv",
    reportNumber = "SINP-TNP-95-11",
    month = "7",
    year = "1995"
}

@article{Dalui:2018,
    author = "Dalui, Surojit and Majhi, Bibhas Ranjan and Mishra, Pankaj",
    title = "{Presence of horizon makes particle motion chaotic}",
    eprint = "1803.06527",
    archivePrefix = "arXiv",
    primaryClass = "gr-qc",
    journal = "Phys. Lett. B",
    volume = "788",
    pages = "486--493",
    year = "2019"
}

@article{Dalui:2019,
    author = "Dalui, Surojit and Majhi, Bibhas Ranjan and Mishra, Pankaj",
    title = "{Induction of chaotic fluctuations in particle dynamics in a uniformly accelerated frame}",
    eprint = "1904.11760",
    archivePrefix = "arXiv",
    primaryClass = "gr-qc",
    journal = "Int. J. Mod. Phys. A",
    volume = "35",
    number = "18",
    pages = "2050081",
    year = "2020"
}

@article{Bombelli:1991eg,
    author = "Bombelli, Luca and Calzetta, Esteban",
    title = "{Chaos around a black hole}",
    reportNumber = "GTCRG-91-12",
    journal = "Class. Quant. Grav.",
    volume = "9",
    pages = "2573--2599",
    year = "1992"
}

@article{Sota:1995ms,
    author = "Sota, Yasuhide and Suzuki, Shingo and Maeda, Kei-ichi",
    title = "{Chaos in static axisymmetric space-times. 1: Vacuum case}",
    eprint = "gr-qc/9505036",
    archivePrefix = "arXiv",
    reportNumber = "WU-AP-45-95",
    journal = "Class. Quant. Grav.",
    volume = "13",
    pages = "1241--1260",
    year = "1996"
}

@article{Vieira:1996zf,
    author = "Vieira, Werner M. and Letelier, Patricio S.",
    title = "{Chaos around a Henon-Heiles inspired exact perturbation of a black hole}",
    eprint = "gr-qc/9604037",
    archivePrefix = "arXiv",
    journal = "Phys. Rev. Lett.",
    volume = "76",
    pages = "1409--1412",
    year = "1996"
}

@article{Suzuki:1996gm,
    author = "Suzuki, Shingo and Maeda, Kei-ichi",
    title = "{Chaos in Schwarzschild space-time: The motion of a spinning particle}",
    eprint = "gr-qc/9604020",
    archivePrefix = "arXiv",
    reportNumber = "WU-AP-59-96",
    journal = "Phys. Rev. D",
    volume = "55",
    pages = "4848--4859",
    year = "1997"
}

@article{Cornish:1996ri,
    author = "Cornish, Neil J. and Frankel, Norman E.",
    title = "{The Black hole and the pea}",
    reportNumber = "UM-P-96-45-REV, UM-P-96-45",
    journal = "Phys. Rev. D",
    volume = "56",
    pages = "1903--1907",
    year = "1997"
}

@article{deMoura:1999wf,
    author = "de Moura, Alessandro P. S. and Letelier, Patricio S.",
    title = "{Chaos and fractals in geodesic motions around a nonrotating black hole with an external halo}",
    eprint = "chao-dyn/9910035",
    archivePrefix = "arXiv",
    journal = "Phys. Rev. E",
    volume = "61",
    pages = "6506--6516",
    year = "2000"
}

@article{Hartl:2002ig,
    author = "Hartl, Michael D.",
    title = "{Dynamics of spinning test particles in Kerr space-time}",
    eprint = "gr-qc/0210042",
    archivePrefix = "arXiv",
    journal = "Phys. Rev. D",
    volume = "67",
    pages = "024005",
    year = "2003"
}

@article{Han:2008zzf,
    author = "Han, Wenbiao",
    title = "{Chaos and dynamics of spinning particles in Kerr spacetime}",
    eprint = "1006.2229",
    archivePrefix = "arXiv",
    primaryClass = "gr-qc",
    journal = "Gen. Rel. Grav.",
    volume = "40",
    pages = "1831--1847",
    year = "2008"
}

@article{Takahashi:2008zh,
    author = "Takahashi, Masaaki and Koyama, Hiroko",
    title = "{Chaotic motion of Charged Particles in an Electromagnetic Field Surrounding a Rotating Black Hole}",
    eprint = "0807.0277",
    archivePrefix = "arXiv",
    primaryClass = "astro-ph",
    journal = "Astrophys. J.",
    volume = "693",
    pages = "472--485",
    year = "2009"
}

@article{Li:2018wtz,
    author = "Li, Dan and Wu, Xin",
    title = "{Chaotic motion of neutral and charged particles in a magnetized Ernst-Schwarzschild spacetime}",
    eprint = "1803.02119",
    archivePrefix = "arXiv",
    primaryClass = "gr-qc",
    journal = "Eur. Phys. J. Plus",
    volume = "134",
    number = "3",
    pages = "96",
    year = "2019"
}

@article{Hashimoto:2016dfz,
    author = "Hashimoto, Koji and Tanahashi, Norihiro",
    title = "{Universality in Chaos of Particle Motion near Black Hole Horizon}",
    eprint = "1610.06070",
    archivePrefix = "arXiv",
    primaryClass = "hep-th",
    reportNumber = "OU-HET-911",
    journal = "Phys. Rev. D",
    volume = "95",
    number = "2",
    pages = "024007",
    year = "2017"
}

@article{Giataganas:2021ghs,
    author = "Giataganas, Dimitrios",
    title = "{Chaotic Motion near Black Hole and Cosmological Horizons}",
    eprint = "2112.02081",
    archivePrefix = "arXiv",
    primaryClass = "hep-th",
    doi = "10.1002/prop.202200001",
    journal = "Fortsch. Phys.",
    volume = "70",
    number = "1",
    pages = "2200001",
    year = "2022"
}

@article{Balbinot:2007kr,
    author = "Balbinot, R. and Fabbri, A. and Farese, S. and Parentani, R.",
    title = "{Hawking radiation from extremal and non-extremal black holes}",
    eprint = "0710.0388",
    archivePrefix = "arXiv",
    primaryClass = "hep-th",
    doi = "10.1103/PhysRevD.76.124010",
    journal = "Phys. Rev. D",
    volume = "76",
    pages = "124010",
    year = "2007"
}

@article{Andrianopoli:2013kya,
    author = "Andrianopoli, Laura and D'Auria, Riccardo and Gallerati, Antonio and Trigiante, Mario",
    title = "{Extremal Limits of Rotating Black Holes}",
    eprint = "1303.1756",
    archivePrefix = "arXiv",
    primaryClass = "hep-th",
    doi = "10.1007/JHEP05(2013)071",
    journal = "JHEP",
    volume = "05",
    pages = "071",
    year = "2013"
}

@article{Galli:2011fq,
    author = "Galli, Pietro and Ortin, Tomas and Perz, Jan and Shahbazi, Carlos S.",
    title = "{Non-extremal black holes of N=2, d=4 supergravity}",
    eprint = "1105.3311",
    archivePrefix = "arXiv",
    primaryClass = "hep-th",
    reportNumber = "IFIC-11-21, IFT-UAM-CSIC-11-19",
    doi = "10.1007/JHEP07(2011)041",
    journal = "JHEP",
    volume = "07",
    pages = "041",
    year = "2011"
}

@article{Lemos:2010kw,
    author = "Lemos, Jose P. S. and Zaslavskii, Oleg B.",
    title = "{Entropy of extremal black holes from entropy of quasiblack holes}",
    eprint = "1011.2768",
    archivePrefix = "arXiv",
    primaryClass = "gr-qc",
    doi = "10.1016/j.physletb.2010.11.033",
    journal = "Phys. Lett. B",
    volume = "695",
    pages = "37--40",
    year = "2011"
}

@article{Kiefer:1998rr,
    author = "Kiefer, Claus and Louko, Jorma",
    title = "{Hamiltonian evolution and quantization for extremal black holes}",
    eprint = "gr-qc/9809005",
    archivePrefix = "arXiv",
    reportNumber = "FREIBURG-THEP-98-18",
    doi = "10.1002/(SICI)1521-3889(199901)8:1<67::AID-ANDP67>3.0.CO;2-6",
    journal = "Annalen Phys.",
    volume = "8",
    pages = "67--81",
    year = "1999"
}

@article{Pradhan:2012yx,
    author = "Pradhan, Parthapratim and Majumdar, Parthasarathi",
    title = "{Extremal Limits and Kerr Spacetime}",
    eprint = "1108.2333",
    archivePrefix = "arXiv",
    primaryClass = "gr-qc",
    doi = "10.1140/epjc/s10052-013-2470-2",
    journal = "Eur. Phys. J. C",
    volume = "73",
    number = "6",
    pages = "2470",
    year = "2013"
}

@article{Howard:2013yqq,
    author = "Howard, E M",
    title = "{Geometric aspects of Extremal Kerr black hole entropy}",
    eprint = "1511.00594",
    archivePrefix = "arXiv",
    primaryClass = "gr-qc",
    doi = "10.4236/jmp.2013.43050",
    journal = "J. Mod. Phys.",
    volume = "4",
    pages = "357",
    year = "2013"
}

@article{Carroll:2009maa,
    author = "Carroll, Sean M. and Johnson, Matthew C. and Randall, Lisa",
    title = "{Extremal limits and black hole entropy}",
    eprint = "0901.0931",
    archivePrefix = "arXiv",
    primaryClass = "hep-th",
    reportNumber = "CALT-68-2717",
    doi = "10.1088/1126-6708/2009/11/109",
    journal = "JHEP",
    volume = "11",
    pages = "109",
    year = "2009"
}

@article{Guica:2008mu,
    author = "Guica, Monica and Hartman, Thomas and Song, Wei and Strominger, Andrew",
    title = "{The Kerr/CFT Correspondence}",
    eprint = "0809.4266",
    archivePrefix = "arXiv",
    primaryClass = "hep-th",
    doi = "10.1103/PhysRevD.80.124008",
    journal = "Phys. Rev. D",
    volume = "80",
    pages = "124008",
    year = "2009"
}

@article{Castro:2010fd,
    author = "Castro, Alejandra and Maloney, Alexander and Strominger, Andrew",
    title = "{Hidden Conformal Symmetry of the Kerr Black Hole}",
    eprint = "1004.0996",
    archivePrefix = "arXiv",
    primaryClass = "hep-th",
    doi = "10.1103/PhysRevD.82.024008",
    journal = "Phys. Rev. D",
    volume = "82",
    pages = "024008",
    year = "2010"
}

@article{Lunin:2001jy,
    author = "Lunin, Oleg and Mathur, Samir D.",
    title = "{AdS / CFT duality and the black hole information paradox}",
    eprint = "hep-th/0109154",
    archivePrefix = "arXiv",
    reportNumber = "OHSTPY-HEP-T-01-019",
    doi = "10.1016/S0550-3213(01)00620-4",
    journal = "Nucl. Phys. B",
    volume = "623",
    pages = "342--394",
    year = "2002"
}

@article{Bhattacharya:2019awq,
    author = "Bhattacharya, Krishnakanta and Dey, Sumit and Majhi, Bibhas Ranjan and Samanta, Saurav",
    title = "{General framework to study the extremal phase transition of black holes}",
    eprint = "1903.03434",
    archivePrefix = "arXiv",
    primaryClass = "gr-qc",
    doi = "10.1103/PhysRevD.99.124047",
    journal = "Phys. Rev. D",
    volume = "99",
    number = "12",
    pages = "124047",
    year = "2019"
}

@inproceedings{Kerr:2007dk,
    author = "Kerr, Roy P.",
    title = "{Discovering the Kerr and Kerr-Schild metrics}",
    booktitle = "{Kerr Fest: Black Holes in Astrophysics, General Relativity and Quantum Gravity}",
    eprint = "0706.1109",
    archivePrefix = "arXiv",
    primaryClass = "gr-qc",
    month = "6",
    year = "2007"
}

@article{Teukolsky:2014vca,
    author = "Teukolsky, Saul A.",
    title = "{The Kerr Metric}",
    eprint = "1410.2130",
    archivePrefix = "arXiv",
    primaryClass = "gr-qc",
    doi = "10.1088/0264-9381/32/12/124006",
    journal = "Class. Quant. Grav.",
    volume = "32",
    number = "12",
    pages = "124006",
    year = "2015"
}

@inproceedings{Visser:2007fj,
    author = "Visser, Matt",
    title = "{The Kerr spacetime: A Brief introduction}",
    booktitle = "{Kerr Fest: Black Holes in Astrophysics, General Relativity and Quantum Gravity}",
    eprint = "0706.0622",
    archivePrefix = "arXiv",
    primaryClass = "gr-qc",
    month = "6",
    year = "2007"
}

@article{Gwak:2022xje,
    author = "Gwak, Bogeun and Kan, Naoto and Lee, Bum-Hoon and Lee, Hocheol",
    title = "{Violation of bound on chaos for charged probe in Kerr-Newman-AdS black hole}",
    eprint = "2203.07298",
    archivePrefix = "arXiv",
    primaryClass = "gr-qc",
    doi = "10.1007/JHEP09(2022)026",
    journal = "JHEP",
    volume = "09",
    pages = "026",
    year = "2022"
}

@article{Jeong:2023hom,
    author = "Jeong, Soyeon and Lee, Bum-Hoon and Lee, Hocheol and Lee, Wonwoo",
    title = "{Homoclinic orbit and the violation of the chaos bound around a black hole with anisotropic matter fields}",
    eprint = "2301.12198",
    archivePrefix = "arXiv",
    primaryClass = "gr-qc",
    reportNumber = "CQUeST-2023-0718",
    doi = "10.1103/PhysRevD.107.104037",
    journal = "Phys. Rev. D",
    volume = "107",
    number = "10",
    pages = "104037",
    year = "2023"
}

@article{Kim:2019hfp,
    author = "Kim, Hyeong-Chan and Lee, Bum-Hoon and Lee, Wonwoo and Lee, Youngone",
    title = "{Rotating black holes with an anisotropic matter field}",
    eprint = "1912.09709",
    archivePrefix = "arXiv",
    primaryClass = "gr-qc",
    doi = "10.1103/PhysRevD.101.064067",
    journal = "Phys. Rev. D",
    volume = "101",
    number = "6",
    pages = "064067",
    year = "2020"
}

@article{Kim:2021vlk,
    author = "Kim, Hyeong-Chan and Lee, Bum-Hoon and Lee, Wonwoo and Lee, Youngone",
    title = "{Charged rotating black hole with an anisotropic matter field: Solution of the Maxwell field}",
    eprint = "2112.04131",
    archivePrefix = "arXiv",
    primaryClass = "gr-qc",
    doi = "10.1063/5.0216304",
    journal = "AIP Conf. Proc.",
    volume = "2874",
    number = "1",
    pages = "020008",
    year = "2024"
}

@article{Kim:2025sdj,
    author = "Kim, Hyeong-Chan and Lee, Wonwoo",
    title = "{Dressing rotating black holes with anisotropic matter}",
    eprint = "2503.06961",
    archivePrefix = "arXiv",
    primaryClass = "gr-qc",
    reportNumber = "CQUeST-2025-0755",
    doi = "10.1140/epjc/s10052-025-15008-w",
    journal = "Eur. Phys. J. C",
    volume = "85",
    number = "11",
    pages = "1245",
    year = "2025"
}

\bibliographystyle{utphys1}

\end{document}